\documentclass[nofootinbib,reprint,onecolumn,amsmath,amssymb,notitlepage,aps,10pt]{revtex4-1}
\usepackage{graphicx}
\usepackage[T1]{fontenc}
\usepackage{dcolumn}
\usepackage{bm}

\usepackage{hyperref}
\hypersetup{colorlinks, linkcolor={NavyBlue},citecolor={NavyBlue},urlcolor={NavyBlue}}

\usepackage[dvipsnames, svgnames, x11names]{xcolor}
\usepackage{float}

\begin{document}
%\preprint{APS/123-QED}

\title{Charged black-bounce spacetimes: Photon rings, shadows and observational appearances}

\author{Yang Guo}\email{ guoy@mail.nankai.edu.cn}
\author{Yan-Gang Miao}\email{Corresponding author: miaoyg@nankai.edu.cn}
%\vspace{0.5cm}

\affiliation{School of Physics, Nankai University, Tianjin 300071, China}

%\date{\today}

\begin{abstract}

The photon ring, shadow and observational appearance of the emission originating near a charged black-bounce are investigated. Based on the geodesic analysis, we determine the upper and lower limits of critical impact parameters of a charged black-bounce. In particular, we find that the charged black-bounce shares the same critical impact parameter with the Reissner-Nordst\"om black hole. In addition, we classify the light  trajectories coming from the region near the charged black-bounce by utilizing  the rays tracing procedure, and then investigate the observational appearance of the emissions from a thin disk accretion and a spherically symmetric infalling accretion.  We reveal that a large charge increases the  observed intensity but decreases the apparent size of shadows, and that the photon ring presents the intrinsic property of a spacetime geometry, which is independent of the types of the two accretions. Our results are in good agreement with the recent observations.

\end{abstract}

\maketitle
\small \tableofcontents
\section{Introduction}
\label{sec:intro}

The studies related to black hole shadows have a long history if we trace their research trajectory. The earliest study was the deflection of light in a strong gravitational field, which was confirmed by the observation of the solar eclipse in 1919 and the subsequent developments associated with the gravitational lensing~\cite{Gott:1984ef,Blandford:1991xc,Bartelmann:1999yn,Lewis:2006fu,Bartelmann:2010fz}. The term ``shadow'' is currently the most common word but it was once called by several interesting names, such as the escape cone~\cite{Synge:1966okc}, the apparent boundary~\cite{Bardeen:1973tla}, and the critical curve~\cite{Gralla:2019xty}, etc., see Ref.~\cite{Perlick:2021aok} for a more detailed review.  In 2000 it was proposed~\cite{Falcke:1999pj} that the shadow of Sagittarius A$^*$ with a thin accretion flow could be observable at submillimeter wavelengths and expected to image event horizons in the future. About two decades later, the  image of the supermassive black hole located at the center of the Messier 87 galaxy (M87$^*$) was announced~\cite{EventHorizonTelescope:2019dse,EventHorizonTelescope:2019uob,EventHorizonTelescope:2019jan,EventHorizonTelescope:2019ths,EventHorizonTelescope:2019pgp,EventHorizonTelescope:2019ggy}  and further the first polarized images of this black hole were released~\cite{EventHorizonTelescope:2021btj} by the Event Horizon Telescope (EHT). These achievements are inspiring great attentions in studying \cite{Bozza:2010xqn,Chen:2021gwy,Rahaman:2021web,Jafarzade:2020ova,Guo:2020nci,Bambhaniya:2021ugr,Wei:2013kza,Zhang:2021hit,Gan:2021xdl,Cunha:2019hzj,Peng:2020wun,Lima:2021las,Peng:2021osd,Churilova:2021tgn} various aspects of the shadow and observational appearance of  black holes.

One of the M87$^*$ features that could be observed is that a thick band of light outlines a dark area, that is, the shadow. Based on the  general relativistic magnetohydrodynamics (GRMHD) simulations, the appearance of the M87$^*$ describes~\cite{EventHorizonTelescope:2019dse} the turbulent, thermal, and magnetized disks orbiting the Kerr black hole. Furthermore, a black hole is expected to show its shadow caused by the gravitational light bending and photon capture at its event horizon
when there are transparent emissions near the black hole. The shadow radius is related to the photon ring which is a geometric property of a spacetime.  It is natural to expect that the photon ring should not depend on the accretion surrounding  the black hole but be determined by the black hole geometry, and that one should observe the same photon ring under various accretion models.

Recently,  a regularizing procedure, i.e. the introduction of a length scale or bounce parameter has been applied~\cite{Franzin:2021vnj} to the Reissner-Nordstr\"om  (RN) black hole in order to generate a candidate spacetime labeled ``charged black-bounce'', ``black-bounce-Reissner-Nordstr\"om geometry'', or ``Reissner-Nordstr\"om-Simpson-Visser (RN-SV) spacetime''. This geometry has three properties: 

(i) It is globally free from curvature singularities;

(ii) It passes all weak-field observational tests;

(iii) It smoothly interpolates between regular black holes and charged traversable wormholes.
 
\noindent
Alternatively, the charged black-bounce can be constructed when an electromagnetic charge is introduced to a family of candidate spacetimes labeled ``black-bounce''~\cite{Simpson:2018tsi,Simpson:2019cer,Lobo:2020kxn,Lobo:2020ffi,Mazza:2021rgq}, where the black-bounce can be derived from the Schwarzschild black hole in terms of the regularizing procedure.
%Besides it has the third property owned by the charged black-bounce, the black-bounce family shares~\cite{Guerrero:2021ues} the same critical impact parameter with the Schwarzschild black hole. Because critical impact parameters describe the intrinsic properties of black-bounce geometry, our concern is whether the introduction of an  electromagnetic charge into a black-bounce remains the critical impact parameter  unchanged. 
The black-bounce without charge interpolating between regular black holes and traversable wormholes shares~\cite{Guerrero:2021ues} the same critical impact parameter with the Schwarzschild black hole because a length scale or bounce parameter in the black-bounce family does not change the critical impact parameter. We expect that the charged black-bounce can share the same critical impact parameter with the RN black hole, that is, the regularizing procedure does not alter the critical impact parameter whether it is applied to the Schwarzschild black hole or the RN black hole.
Moreover, we discuss the photon rings and shadows of charged black-bounces due to their close association with~\cite{Okyay:2021nnh,Bronnikov:2021liv} critical impact parameters.

In this paper, we attempt to provide an investigation to the observational appearance of charged black-bounces, focusing on photon rings and shadows. We show the images of the appearance of charged black-bounces under various illumination conditions. These images  can help us to  test  the strong gravitational  field around a compact object described by the charged black-bounce.

The outline of this paper is as follows. In Sec.~\ref{sec:geometry} we review briefly the  properties of a charged black-bounce geometry and then determine the boundaries of critical impact parameters. Next,  we trace the light rays coming from the region near a charged black-bounce for the classification of light rays in Sec.~\ref{sec:light bending}.  We investigate the appearance of the region near a charged black-bounce when emissions come from a thin disk accretion and a spherically symmetric infalling accretion in Sec.~\ref{sec:appearance}. Our results are discussed and compared with astronomical observations in Sec.~\ref{sec:compare}. Finally, we give our  concluding remarks in Sec.~\ref{sec:con}.

\section{The field sources and null geodesics of Reissner-Nordst\"om-Simpson-Visser spacetimes}
\label{sec:geometry}

\subsection{Field sources for Einstein's equations}
We start with the ``regularizing procedure'' recently proposed in Ref.~\cite{Franzin:2021vnj}, in which the radial coordinate $r$ in the RN metric is replaced by $\sqrt{x^2+a^2}$, where the parameter $a$ is associated with the Plank length. The charged black-bounce is a one-parameter modification of the  Reissner-Nordst\"om
black hole of General Relativity, and it is called black-bounce-Reissner-Nordst\"om (BB-RN), charged black-bounce spacetime, or Reissner-Nordst\"om-Simpson-Visser (RN-SV) spacetime. It can be obtained as an exact solution to the Einstein equation sourced by a combination of a minimally coupled phantom scalar field  and a  nonlinear electrodynamics field. The action reads \cite{Bronnikov:2021uta},
\begin{eqnarray}
	S=\int\sqrt{-g}\,d^4x \left( \mathcal{R} + 2 \epsilon \partial_\mu\phi \partial_\nu\phi - 2V(\phi) 
	- \mathcal{L(F)} \right),\label{action}
\end{eqnarray}
where $\mathcal{L(F)}$ is the gauge-invariant Lagrangian density of nonlinear electrodynamics with $\mathcal{F}\equiv F_{\mu\nu}F^{\mu\nu}$ and  $\epsilon=-1$ for a phantom scalar field.  The Lagrangian density and the potential of an uncharged scalar field $\phi(x)$ take the following forms,
\begin{eqnarray}
	\mathcal{L(F)} 	= \frac{12 Ma^2}{5 (2q^2/\mathcal{F})^{5/4}} 
	+ \frac {2Q^2 \big[3(2q^2/\mathcal{F})^{1/2}-4a^2\big]}{3(2q^2/\mathcal{F})^{3/2}},			
	\label{V-fin}   
\end{eqnarray}
and
\begin{eqnarray}
	V(\phi) = \frac {2 \cos^6 \phi}{15 a^4} (6M a \sec \phi - 5Q^2),
\end{eqnarray}
where $M$ is mass of the black-bounce, $q$ is magnetic charge of free nonlinear electrodynamics and $Q$ electric charge parameter. Varying Eq.~(\ref{action}) with respect to the metric yields Einstein's equation,
\begin{eqnarray}
	G_{\mu\nu}=T_{\mu\nu},
\end{eqnarray}
where the energy-momentum tensor, $T_{\mu\nu}=-T_{\mu\nu}[\phi]-T_{\mu\nu}[\mathcal{F}]$, is the  combination of the energy-momentum tensor of scalar field and the energy-momentum tensor of nonlinear electromagnetic field. One can obtain the RN-SV metric by following the standard procedure in the four dimensional spacetime,
\begin{eqnarray}
	ds^2=-A(x)dt^2+A^{-1}(x)dx^2+r^2(x)d\Omega^2, \qquad r(x)\equiv\sqrt{x^2+a^2},
\end{eqnarray}
with
\begin{eqnarray}
         A(x)=1-\frac{2M}{r(x)}+\frac{Q^2}{r^2(x)}.
\end{eqnarray}

The radial coordinate expands to the entire real domain,  $x\in(-\infty,+\infty)$, and $d\Omega^2$ is  the line element of a unit 2-sphere. This procedure is a smooth transformation\footnote{Note that this procedure is not a coordinate transformation~\cite{Simpson:2021vxo}, which leaves $dx$ undisturbed and makes the metric components an explicit $x$-dependence.} which makes it evident that the charged black-bounce is a globally regular spacetime. The charged black-bounce will recover the RN geometry when the length scale $a$ approaches zero, whilst it will recover~\cite{Morris:1988tu,Boonserm:2018orb} the Morris-Thorne wormhole when $M\rightarrow0$ and $Q\rightarrow0$. Furthermore, its horizons are located at
 \begin{eqnarray}
 	x_{\rm h}=S_1\sqrt{\left( M+S_2\sqrt{M^2-Q^2}\right) ^2-a^2},
 \end{eqnarray}
where $S_1,S_2=\pm1$, $S_1=1$ corresponds to our universe while $S_1=-1$ the copy of our universe, and $S_2 =1$ indicates outer horizon while $S_2 =-1$ inner horizon, respectively. Due to the introduction of the bounce parameter, the geometry and the corresponding structure of event horizons are significantly deformed. The  constraints on the three parameters $(M, Q, a)$ for different geometric types are as follow:
\begin{itemize}
	\item $|Q|< M$ and $a< M\pm\sqrt{M^2-Q^2}$ 
	
	This case corresponds to a charged regular black hole with a standard outer (inner) horizon.
	
	\item $|Q|< M$ and $a= M \pm\sqrt{M^2-Q^2}$,  or  $|Q|= M$ and $a=M$
	
	This case corresponds to a non-traversable wormhole since the geometry possesses an event horizon at the throat.
	
	\item $|Q|< M$ and $a> M \pm\sqrt{M^2-Q^2}$,
	or $|Q|= M$ and $a>M$
	
	In this geometry the inner horizons disappear and then the outer horizons disappear. Thus, this case corresponds to a traversable wormhole.
	
	\item $|Q|=M$ and $ a< M$ 
	
	This case corresponds to an extreme black hole with an extremal horizon at $x_{\rm h}=\sqrt{M^2-a^2}$.
	\item $|Q|>M$ 
	
	There are no horizons in this geometry. This case corresponds to a traversable wormhole.
\end{itemize}

\subsection{Null geodesics and impact parameters}

A freely falling massless particle moving along a null geodesic satisfies the equation,
\begin{eqnarray}
	g_{\mu\nu}\dot{x}^{\mu}\dot{x}^{\nu}=0,
\end{eqnarray}
where the dot stands for the derivative with respect to  affine parameter $\lambda$. Without loss of generality, let us pay attention to the orbits in the equatorial hyperplane ($\theta=\pi/2$). Due to Killing symmetries in this spacetime, one has the conserved energy and conserved angular momentum,
\begin{eqnarray}
E=A(x)\dot{t}, \qquad  L=r^2(x)\dot{\varphi}.\label{conengamom}
\end{eqnarray}
   The impact parameter is defined by the ratio, 
  \begin{eqnarray} 
   b\equiv L/E,\label{imppar} \end{eqnarray} 
   which is only relevant to the trajectory of the null geodesic. By re-parameterizing the affine parameter, we obtain the null geodesic equation as follows,
\begin{eqnarray} \label{eq:geo}
	\dot{x}^2+V_{\rm eff}(x)=\frac{1}{b^2} ,
\end{eqnarray}
with the effective potential,
\begin{eqnarray} \label{eq:Veff}
	V_{\rm eff}(x)=\frac{A(x)}{r^2(x)}.
\end{eqnarray}

\begin{figure}[htbp]
	\centering
	\includegraphics[width=0.6\linewidth]{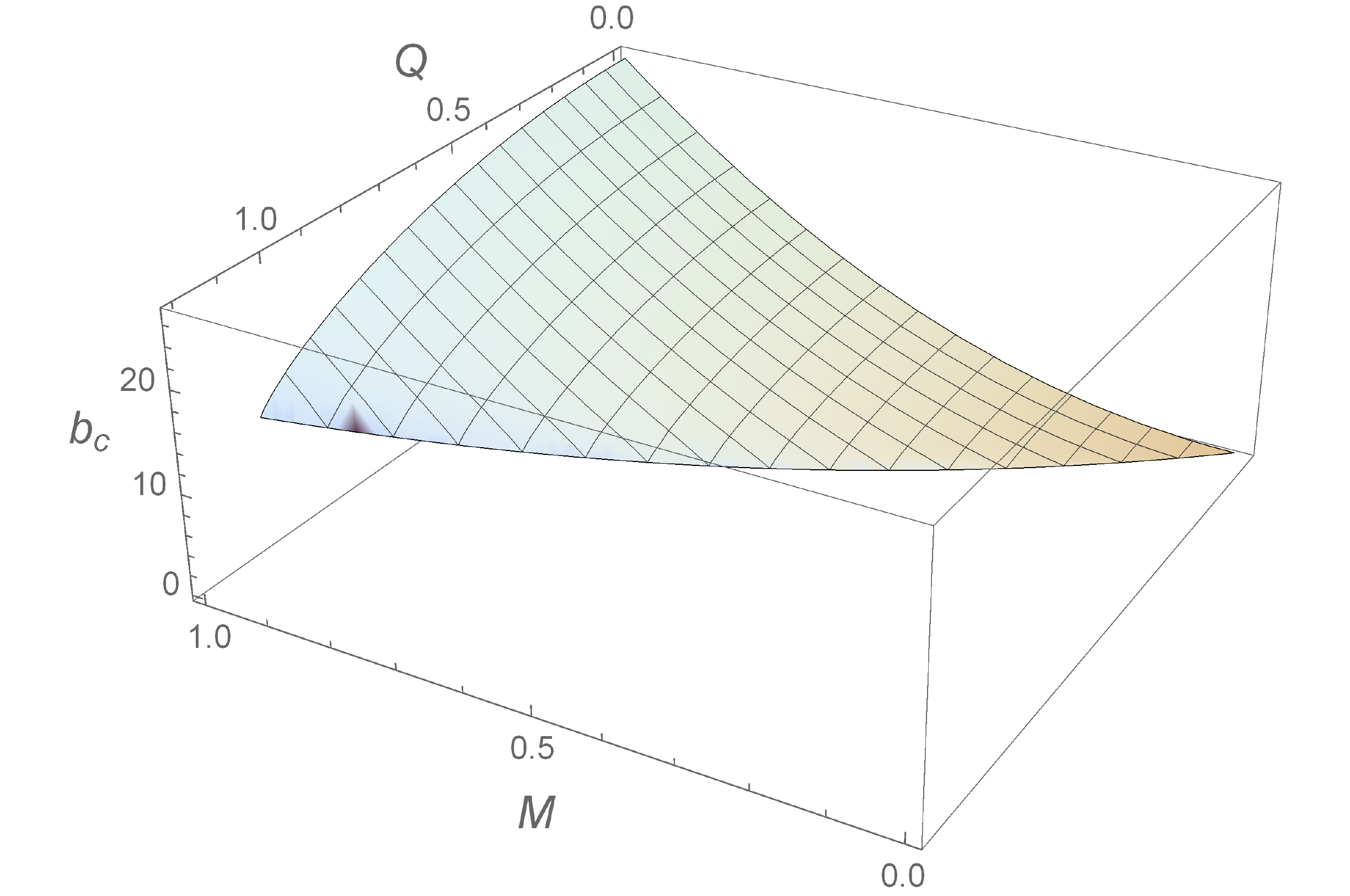}
	\caption{The critical impact parameter as a function of $M$ and $Q$ in a charged black-bounce.} \label{Ipact}
\end{figure}

The particle moving along the geodesics with $b<b_c$ will fall into the event horizon of a charged black-bounce. Here  the critical impact parameter $b_c$ in this  spacetime is charge dependent, that is,
\begin{eqnarray}
	b_c&=&\frac{r(x_{\rm ph})}{\sqrt{A(x_{\rm ph})}} \nonumber\\
	&=& \frac{3 \mathcal{A}^{1/2}+\mathcal{B}}{\left( 9 M^2-8 M  \left( 3 \mathcal{A}^{1/2}+\mathcal{B}\right)^{1/2} +6 \mathcal{A}^{1/2}+\mathcal{B}\right)^{1/2} }, \label{IP}
\end{eqnarray}
where $x_{\rm ph}$ is the photon sphere radius, $\mathcal{A}\equiv 9 M^4-8 M^2 Q^2$, and $\mathcal{B}\equiv 9M^2-4Q^2$. Based on Eq.~(\ref{IP}), we plot in Fig.~\ref{Ipact} the critical impact parameter with respect to the charge and mass. The critical impact parameter decreases continuously with the increasing of charge and reaches its minimum when $|Q|=M$. It is natural to have the maximum critical impact parameter when $|Q|=0$, where this maximum value  equals~\cite{Gralla:2019xty} the critical impact parameter of Schwarzschild black holes. We note that the charged black-bounce shares\footnote{In black-bounce spacetimes, all black-bounce solutions have~\cite{Guerrero:2021ues} the same critical impact parameter as the Schwarzschild black hole has. Here the fact that the length scale $a$ does not appear in Eq.~(\ref{IP}) maintains the critical impact parameter unchanged in RN black holes or charged black-bounces.} the same critical impact parameter with the RN black hole. According to this property, we have the following two boundaries of the critical impact parameter in the charged black-bounce spacetime,
\begin{eqnarray}
	b^{\rm min}_{c}=4M,\label{bcmin}
\end{eqnarray}
\begin{eqnarray}
 b^{\rm max}_{c}=3\sqrt{3}M.\label{bcmax}%\approx5.1962M.
\end{eqnarray}

\section{Rays tracing and light bending near a charged black-bounce}
\label{sec:light bending}

A key procedure we have to follow before we capture the appearance of emissions is to trace the light rays coming from the region near a charged black-bounce. The total number of orbits is defined~\cite{Gralla:2019xty} as $n(b)\equiv \theta/(2\pi)$, where $\theta$ is the total change of azimuth angle. It measures the number of times the null geodesics cross an equatorial plane and can be classified into three types:
    \begin{enumerate}
	\item Direct emission: $n < \frac{3}{4}$, which means a null geodesic intersects the equatorial plane only once;
	\item Lensed: $\frac{3}{4} < n < \frac{5}{4}$, which means a null geodesic intersects the equatorial plane twice;
	\item Photon ring: $n > \frac{5}{4}$, which means a null geodesic intersects the equatorial plane at least three times.
\end{enumerate}

\begin{figure}[htbp]
	\centering
	
\includegraphics[width=0.6\linewidth]{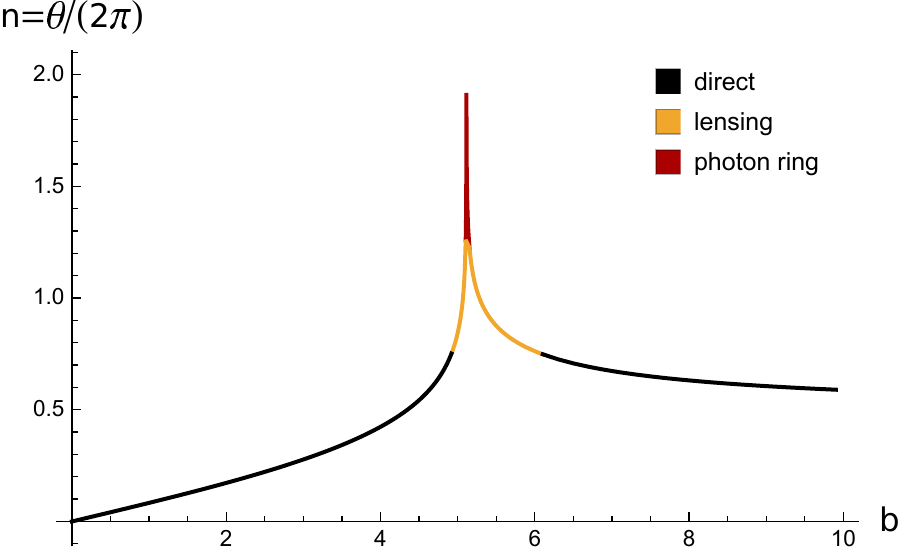}
	
	\caption{The number of orbits as a function of impact parameter $b$. The direct ($n<3/4$), lensed ($3/4<n<5/4$), and photon ring ($n>5/4$) trajectories  are colored in black, gold, and red, respectively.} \label{orbit}
\end{figure}

The three types of light rays are plotted in Fig.~\ref{orbit} with the setting of $M=1$, $Q=0.1$, and $a=0.5$, in which we show the fractional number of orbits as a function of impact parameter $b$. The singularity of the total number appears at $b=b_c\approx 5.1875$, which belongs of course to the range depicted by Eqs.~(\ref{bcmin}) and (\ref{bcmax}), i.e., $b_c\in (4, 3\sqrt{3})$.

\begin{figure}[htbp]
	\centering
	\includegraphics[width=0.6\linewidth]{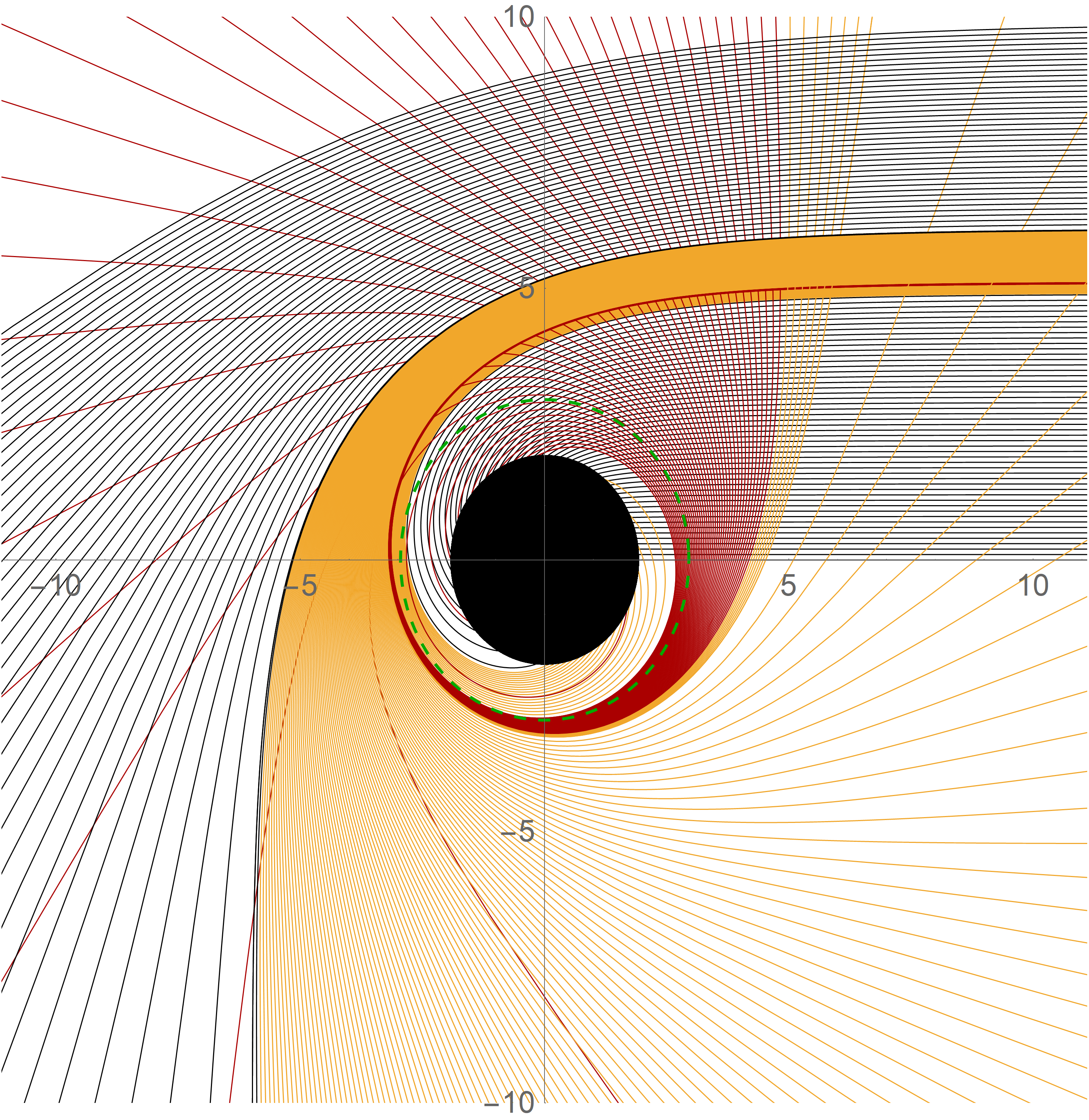}
	\caption{The photon trajectories near the charged black-bounce (shown as a black solid disk) in $(r, \theta)$ Euclidean polar coordinates. The spacing of the impact parameter for plotting is chosen to be 1/10, 1/100, and 1/1000 in the direct (black), lensed (gold), and photon ring (red) bands, respectively.} \label{Rays}
\end{figure}

We plot Fig.~\ref{Rays} in order to give a clearer picture of the photon trajectories for our rays tracing. If the right (East) of Fig.~\ref{Rays} is regarded as the ``north pole direction'' as used in Ref.~\cite{Gralla:2019xty}, we notice that the black curves cross the equatorial plane only once, the gold ones do at least twice, and the red ones do at least three times. In addition, it is worth noting that the light rays with $b=b_c$ spiral towards an unstable circular orbit on the photon sphere, see the green dashed curve around the black solid disk. The light rays with $b<b_c$ fall into the event horizon and only the light rays with $b>b_c$ can be captured by the observer.  In the black-bounce spacetime ($Q=0$) there is another way~\cite{Simpson:2018tsi} that turns its geometry into a wormhole when $a>2M$. In such a wormhole case some rays with $b<b_c$, called~\cite{Guerrero:2021ues} the retro-orbits,  contribute to the luminosity in the observer's screen.

\section{Observational appearances of a charged black-bounce}
\label{sec:appearance}
With all the configurations in place we now focus our attention on the appearance of a charged black-bounce under various illumination conditions. An observer at the north pole in a face-on orientation to the equatorial plane will receive the  isotropic emissions from the accretion disk lying in the equatorial plane. The observed intensity at frequency $\nu^{\prime}$ and the specific intensity of the emission at frequency $\nu$ are given \cite{Gralla:2019xty} by
\begin{eqnarray}
	I^{\rm obs}_{\nu^{\prime}}=g^3I^{\rm em}_{\nu}, \qquad g=[A(x)]^{1/2}.
\end{eqnarray} 
Integrating over all frequencies  yields $I^{\rm obs}=g^4I^{\rm em}$, and  summing all intensities from each intersection with the disk gives the total observed intensity,
\begin{eqnarray}
	I^{\rm obs}(b)=\sum_mg^4I^{\rm em}|_{x=x_m(b)},
\end{eqnarray}
  where $x_m(b)$ is the radial coordinate of the $m$-th intersection with the disk plane  hit by the light ray with impact parameter $b$ and it is also called the {\em transfer function}.

\subsection{Thin disk accretions}
The first model of accretion disks we consider is the emission from the innermost stable circular orbit (ISCO) whose intensity of emission satisfies \cite{Li:2021riw} the following rule,
\begin{eqnarray}
	I^{\rm em}(x) = \begin{cases}
		\Big({x - (x_{\rm isco} - 1)}\Big)^{-2}, &\;\;\; x> x_{\rm isco},\\
		0, & \;\;\; x\leq x_{\rm isco},
	\end{cases}
\end{eqnarray}
where $x_{\rm isco}$ is the radius of the innermost stable circular orbit.

\begin{figure*}[htbp]
	\centering
	\includegraphics[width=5.9cm,height=5.0cm]{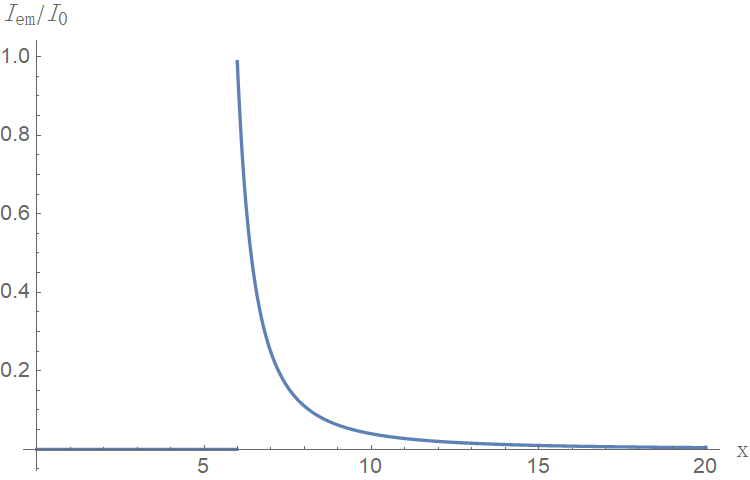}
	\includegraphics[width=5.9cm,height=5.0cm]{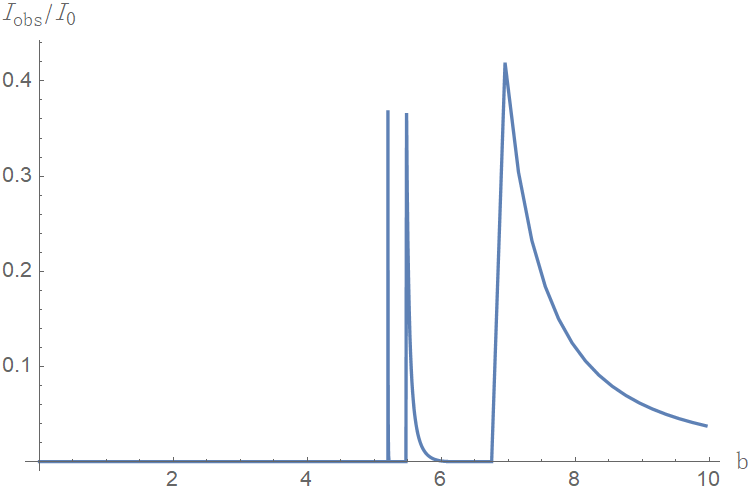}
	\includegraphics[width=5.9cm,height=5.0cm]{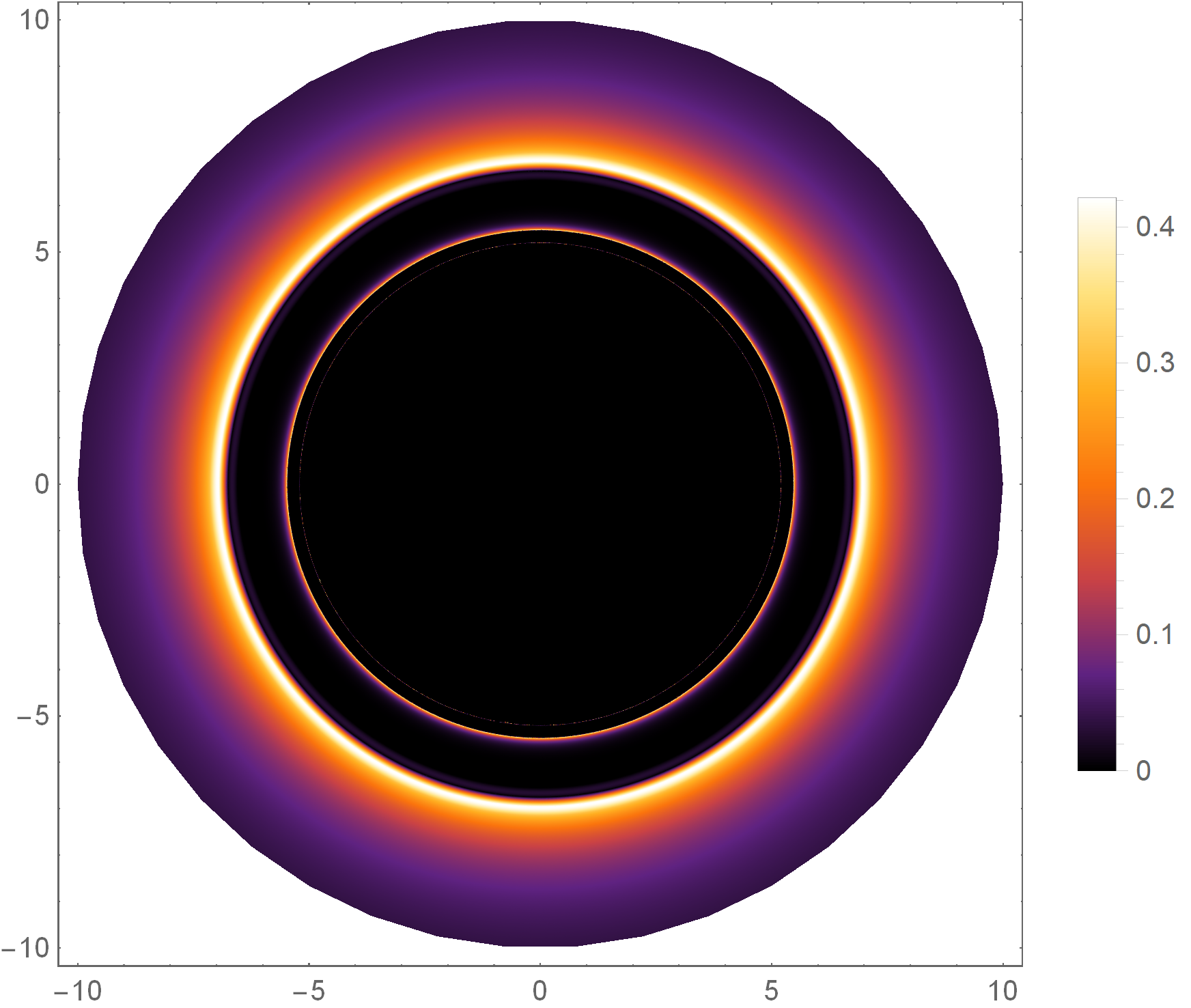}
	\includegraphics[width=5.9cm,height=5.0cm]{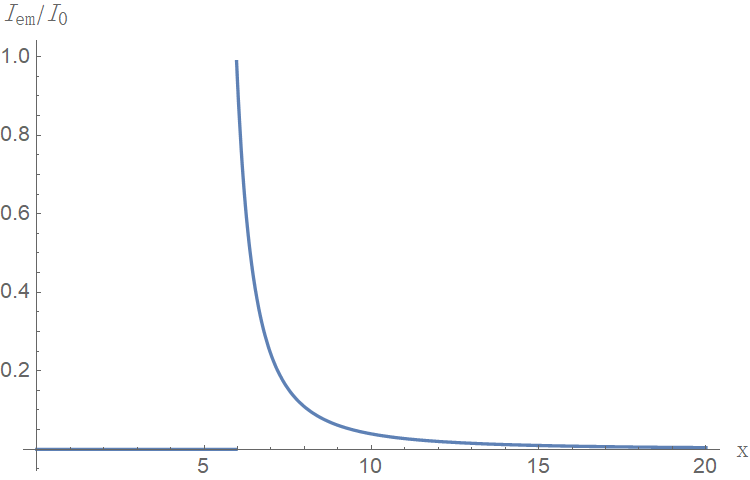}
	\includegraphics[width=5.9cm,height=5.0cm]{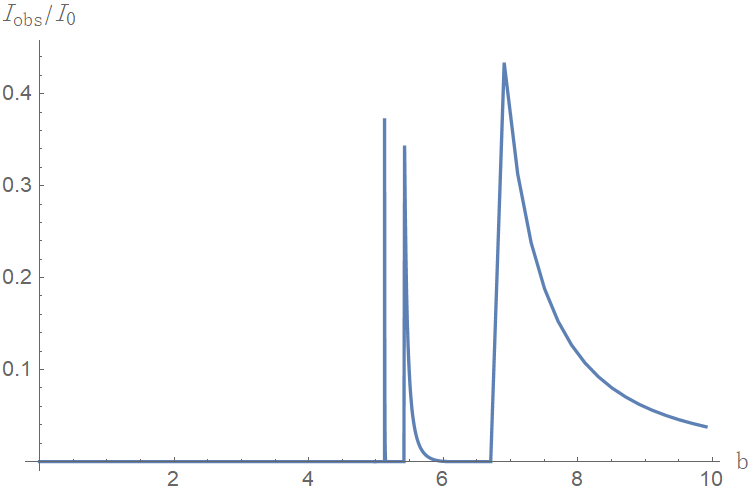}
	\includegraphics[width=5.9cm,height=5.0cm]{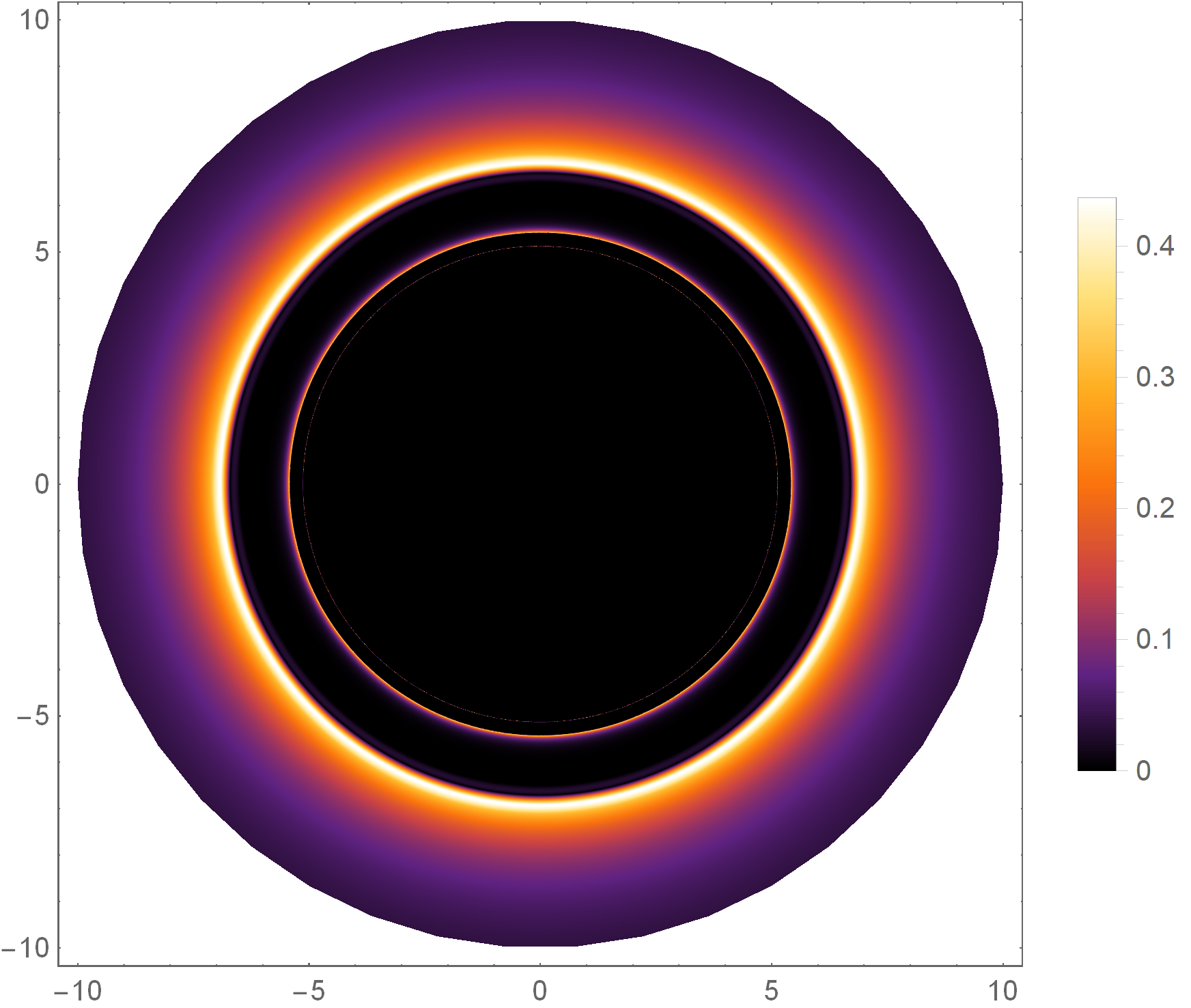}
	\includegraphics[width=5.9cm,height=5.0cm]{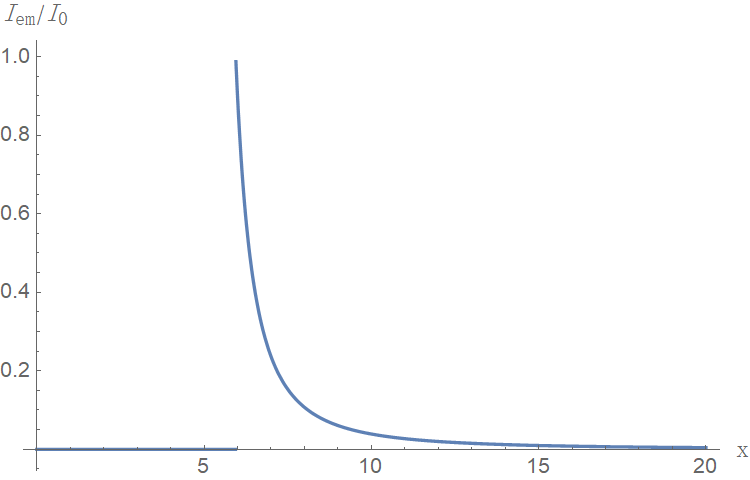}
	\includegraphics[width=5.9cm,height=5.0cm]{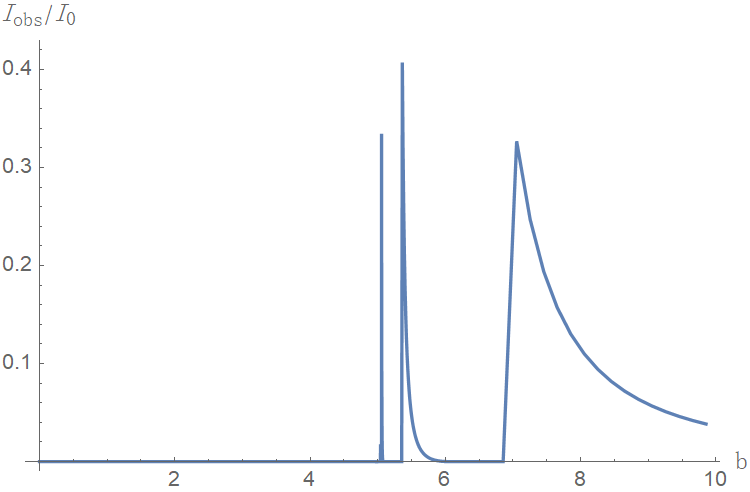}
	\includegraphics[width=5.9cm,height=5.0cm]{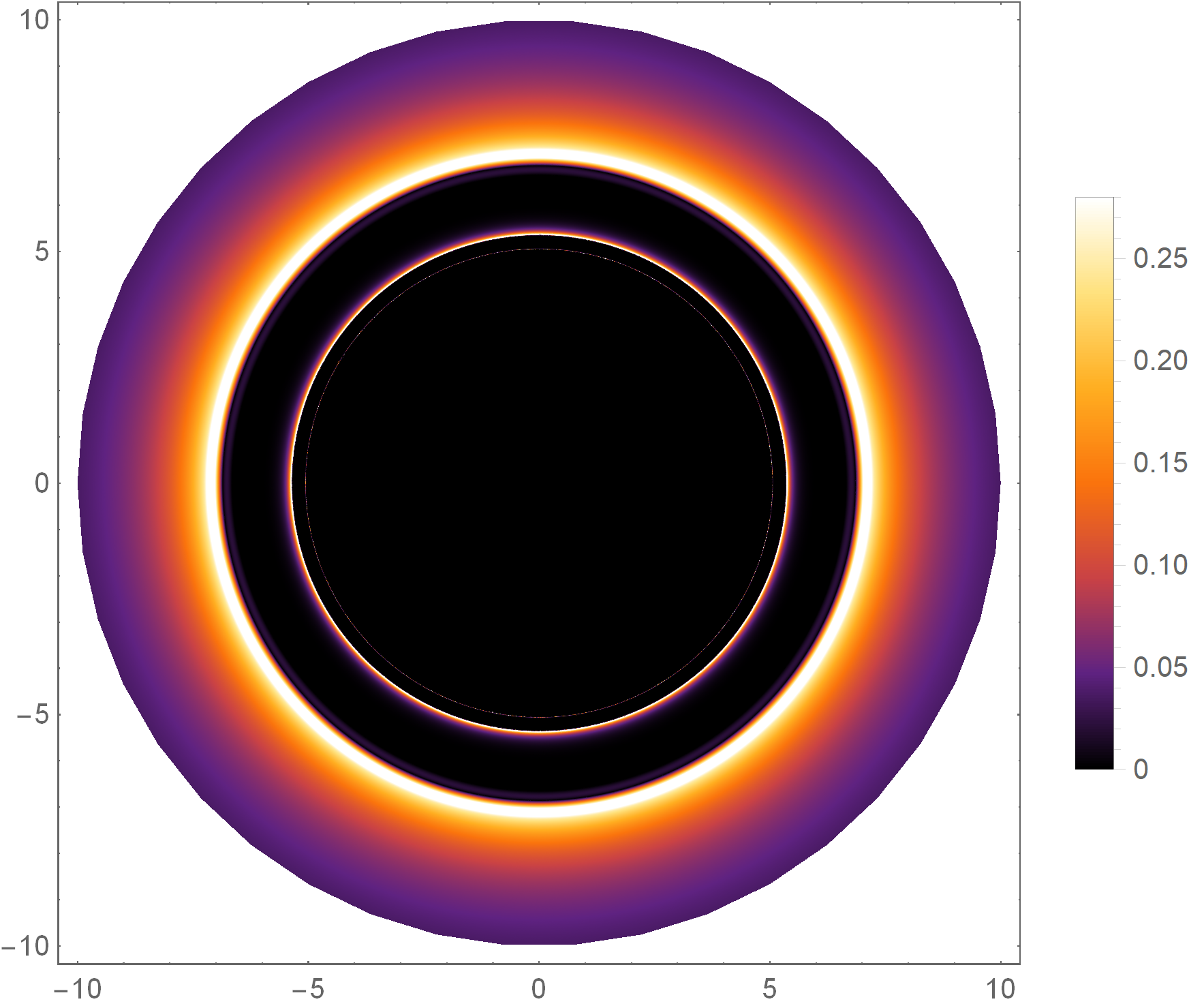}
	\caption{The observational appearance of emission near the charged black-bounce surrounded by a thin disk accretion, where the observer is located at the north pole, facing on the orientation to the equatorial plane. From left to right, the panels show the emitted intensity,  observed intensity, and optical appearance, respectively. From top to bottom, the charge and bounce parameter are set to be $Q=0.1$ and $a=0$ (RN case, top), $Q=0.1$ and $a=0.5$ (middle), and $Q=0.1$ and $a=0.7$ (bottom), respectively.}
	\label{fig:thin}
\end{figure*}
In Fig.~\ref{fig:thin} we show the appearance of the region near a charged black-bounce surrounded by a thin disk accretion. There are two common features in the three cases. One feature is that the gravitational redshift effect does not significantly reduce the observed intensity because the emitted peak is outside the photon orbit in each case and the direct emission dominates the luminosity in the observational appearance. The other feature is that the photon rings have the same size for different values of the bounce parameter, which further confirms that the charged black-bounce has the same critical impact parameter with that of RN black holes. In the top row (RN case with $M=1$, $Q=0.1$, and  $a=0$),  the emitted intensity is sharply peaked near $x\approx5.98$ (see the left panel) which is outside the photon orbit at $x\approx2.99$. The middle row of Fig.~\ref{fig:thin} depicts the charged black-bounce with $M=1$, $Q=0.1$, and $a=0.5$. The emitted intensity peaks at $x\approx5.96$ which is also outside the photon orbit at $x\approx2.95$. And the bottom row of Fig.~\ref{fig:thin} presents the charged black-bounce with $M=1$, $Q=0.1$, and $a=0.7$. The emitted intensity and photon orbit are located at $x\approx5.58$ and $x\approx2.78$, respectively. The contributions from the photon ring and lensing ring to the total observed intensity are negligible because they appear in the innermost layer as a thin and faint ring (see the right panel). Thus, the main contribution to the total luminosity in the observational appearance comes from the direct emission that presents a wide and bright ring.

\begin{figure*}[htbp]
	\centering
	\includegraphics[width=5.9cm,height=5.0cm]{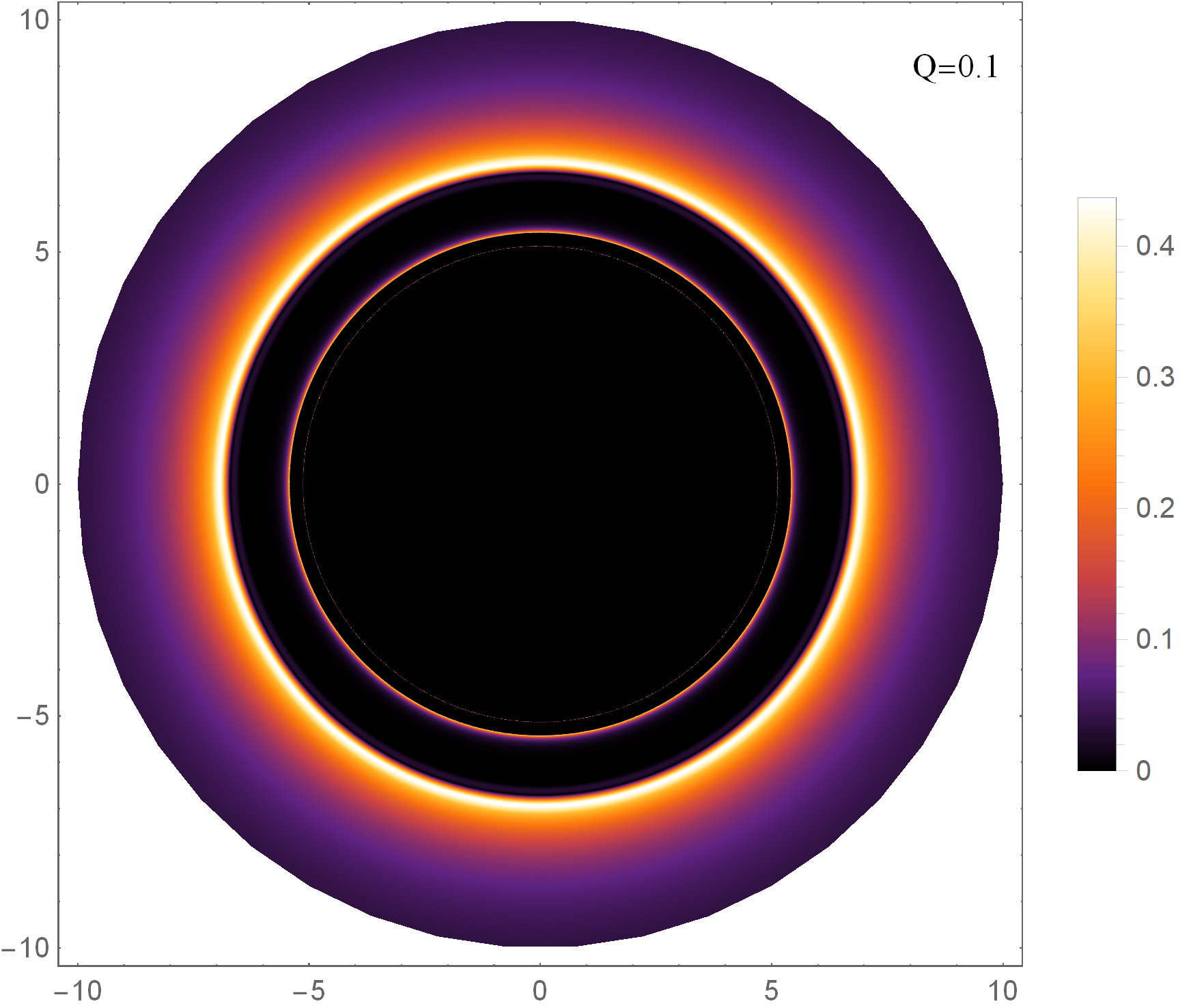}
	\includegraphics[width=5.9cm,height=5.0cm]{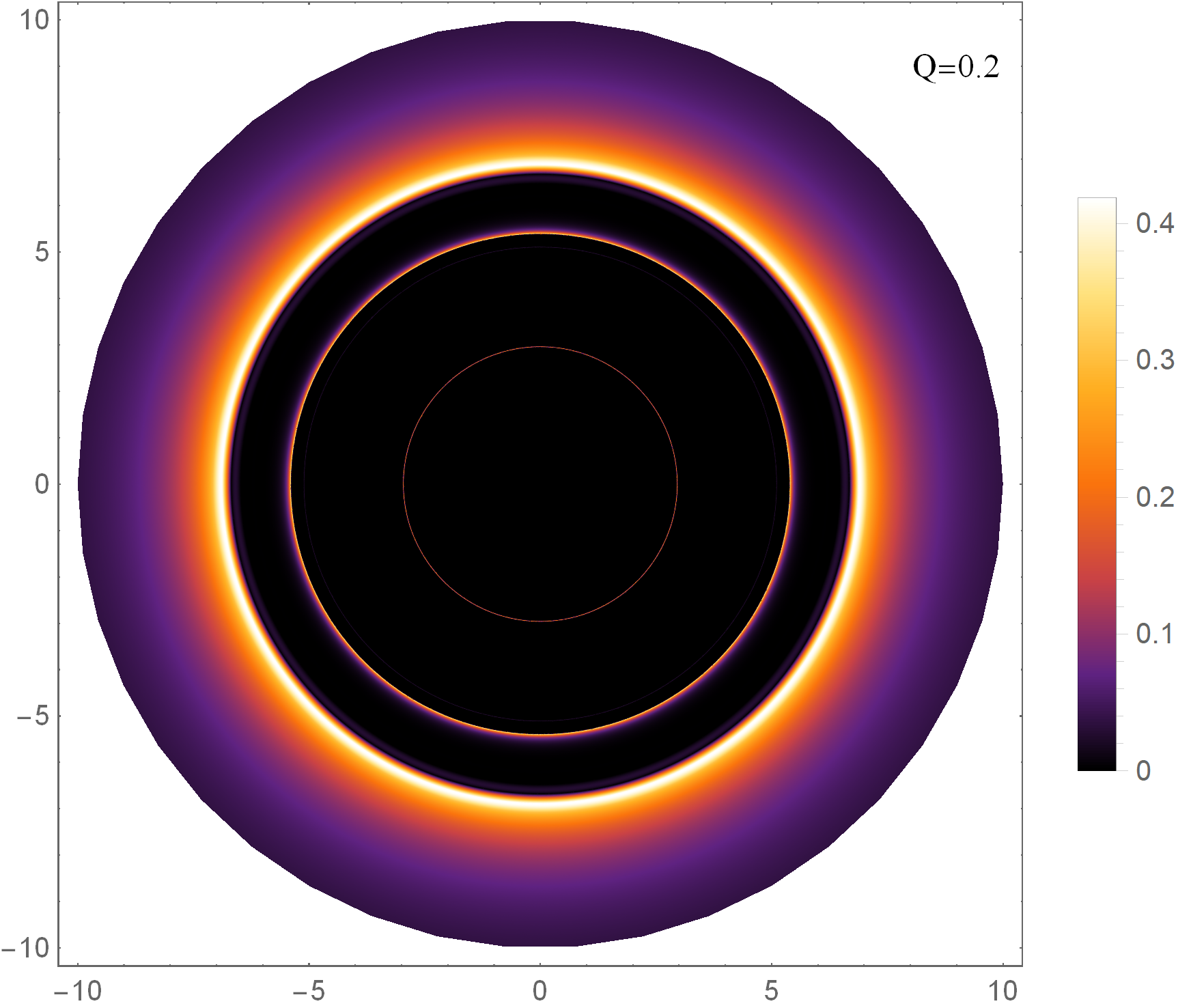}
	\includegraphics[width=5.9cm,height=5.0cm]{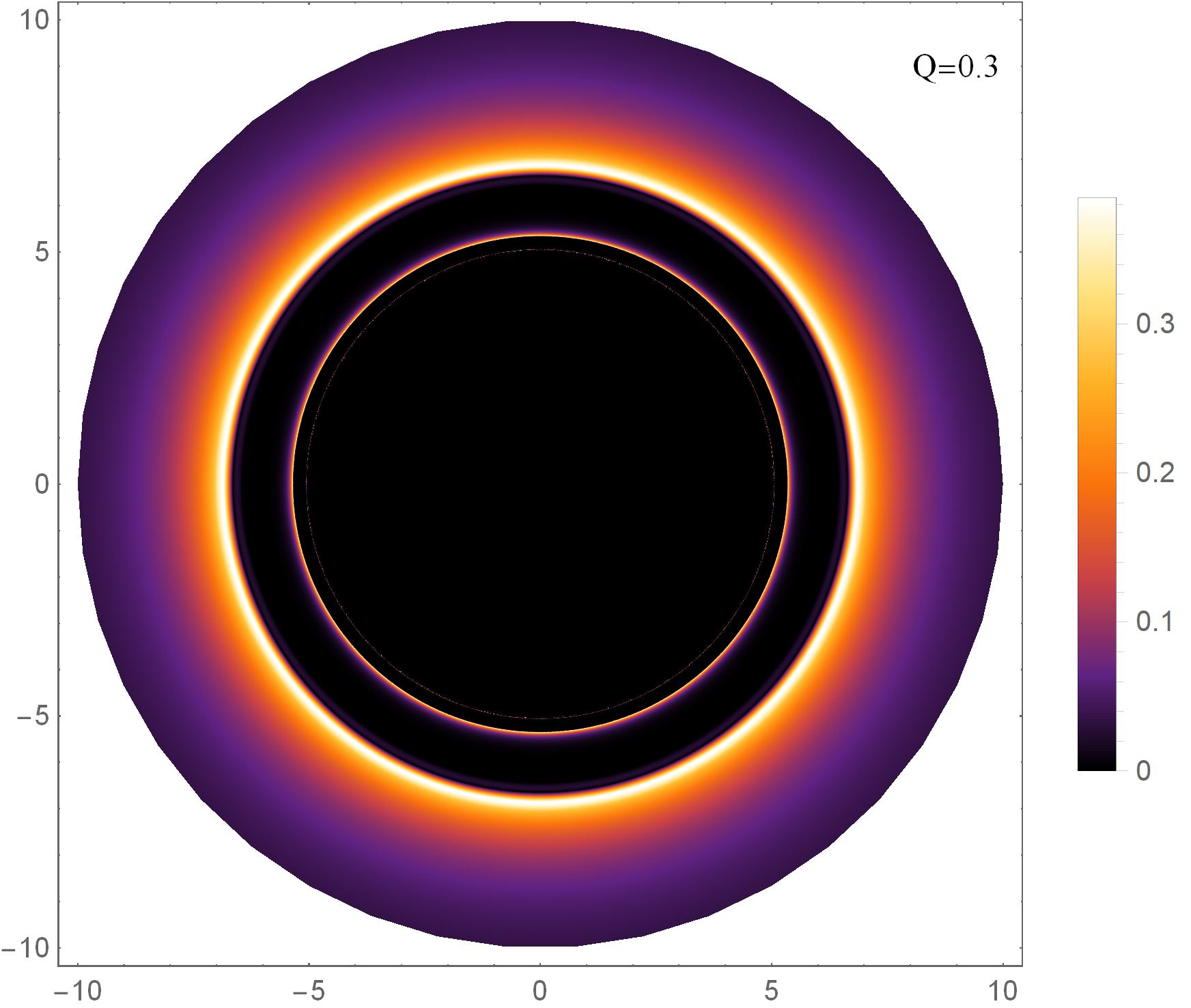}
	\includegraphics[width=5.9cm,height=5.0cm]{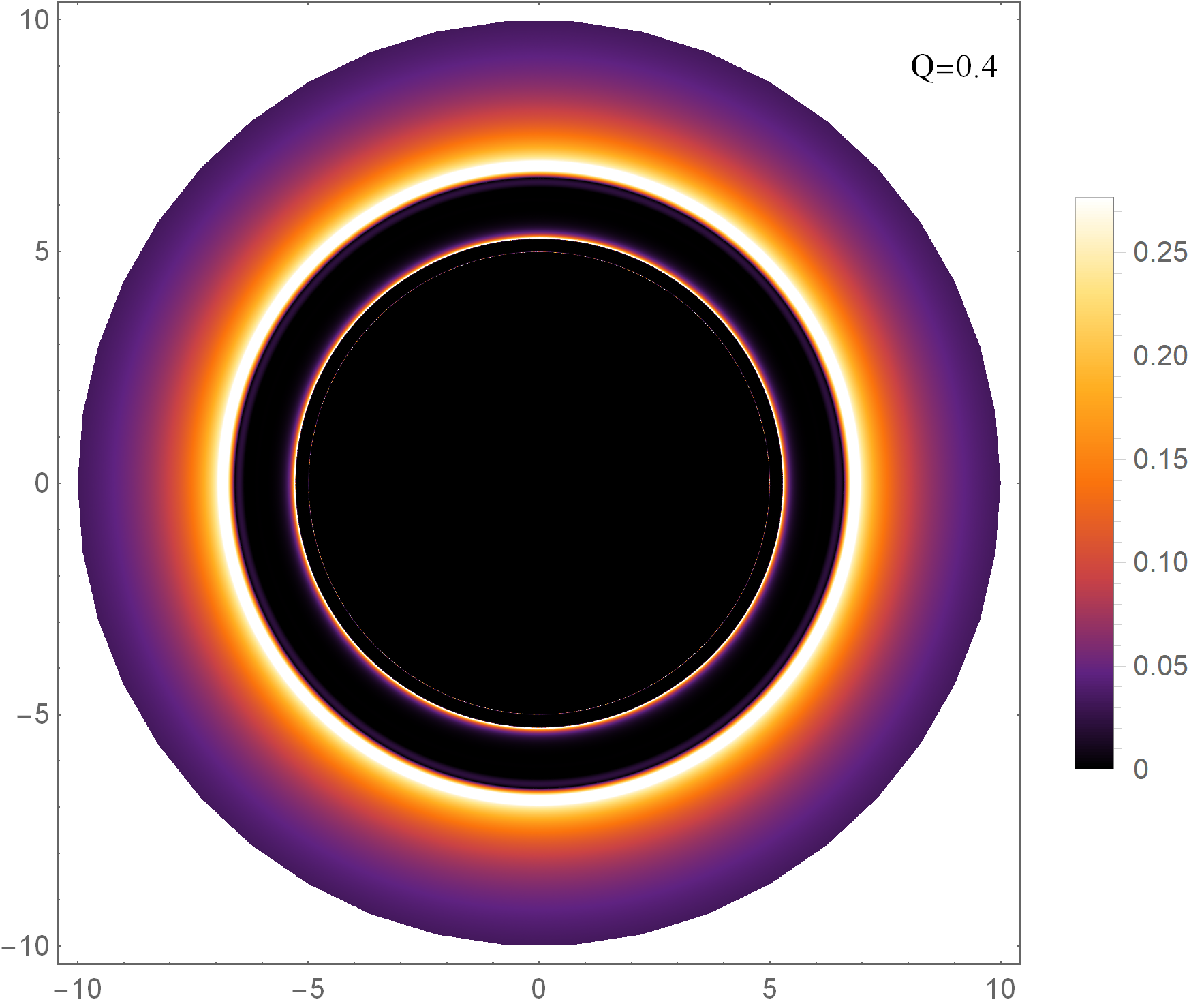}
	\includegraphics[width=5.9cm,height=5.0cm]{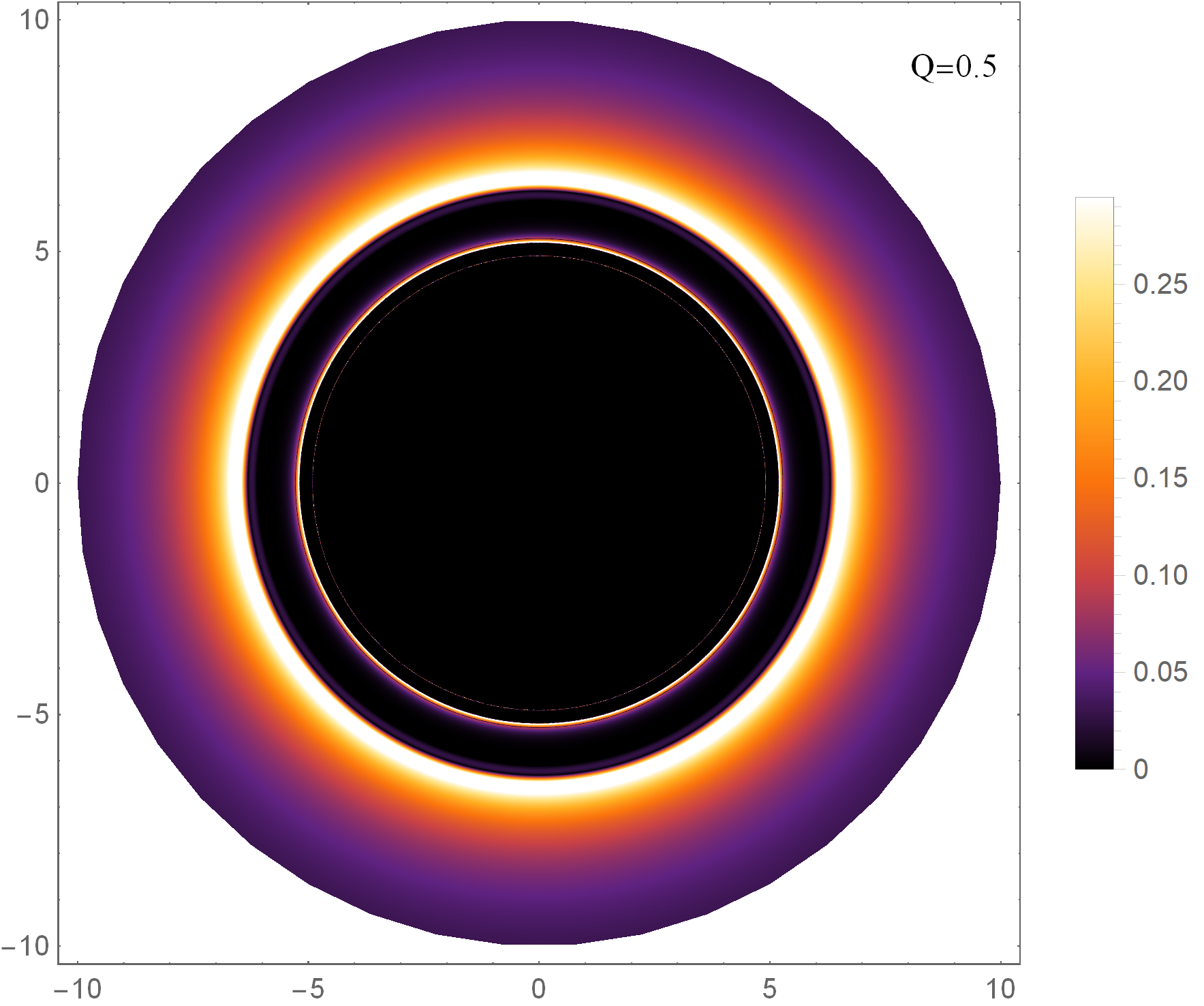}
	\includegraphics[width=5.9cm,height=5.0cm]{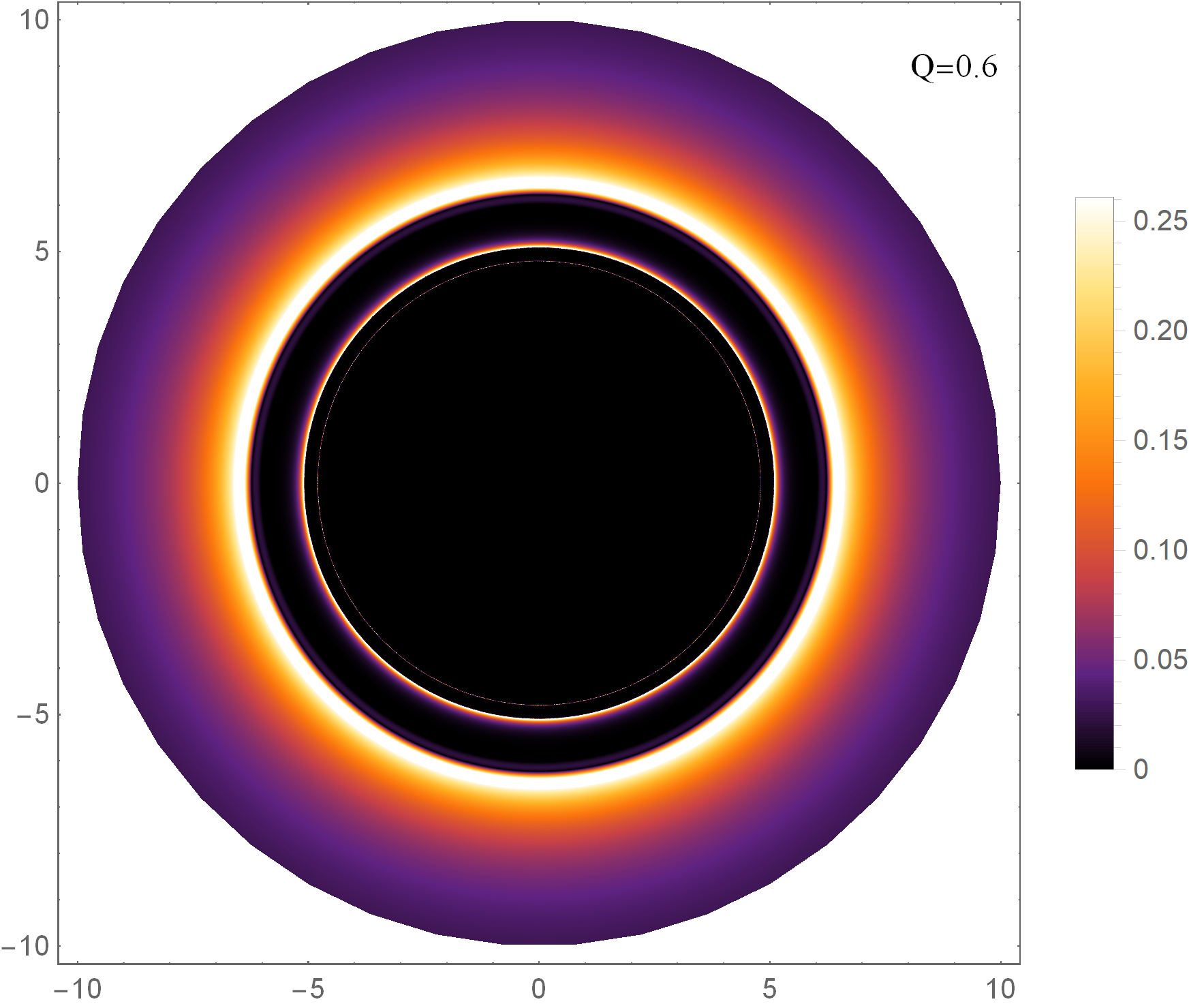}
	\includegraphics[width=5.9cm,height=5.0cm]{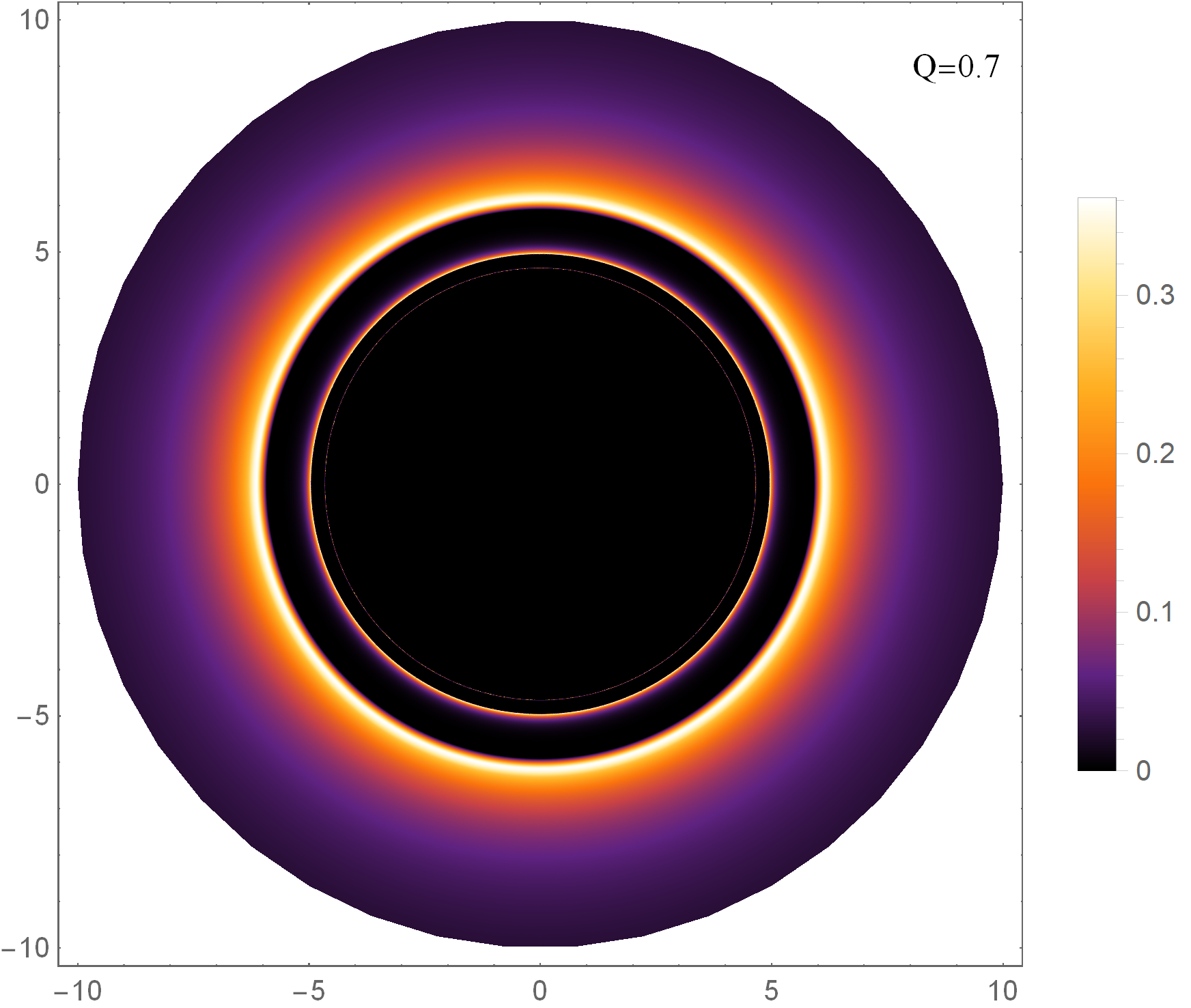}
	\includegraphics[width=5.9cm,height=5.0cm]{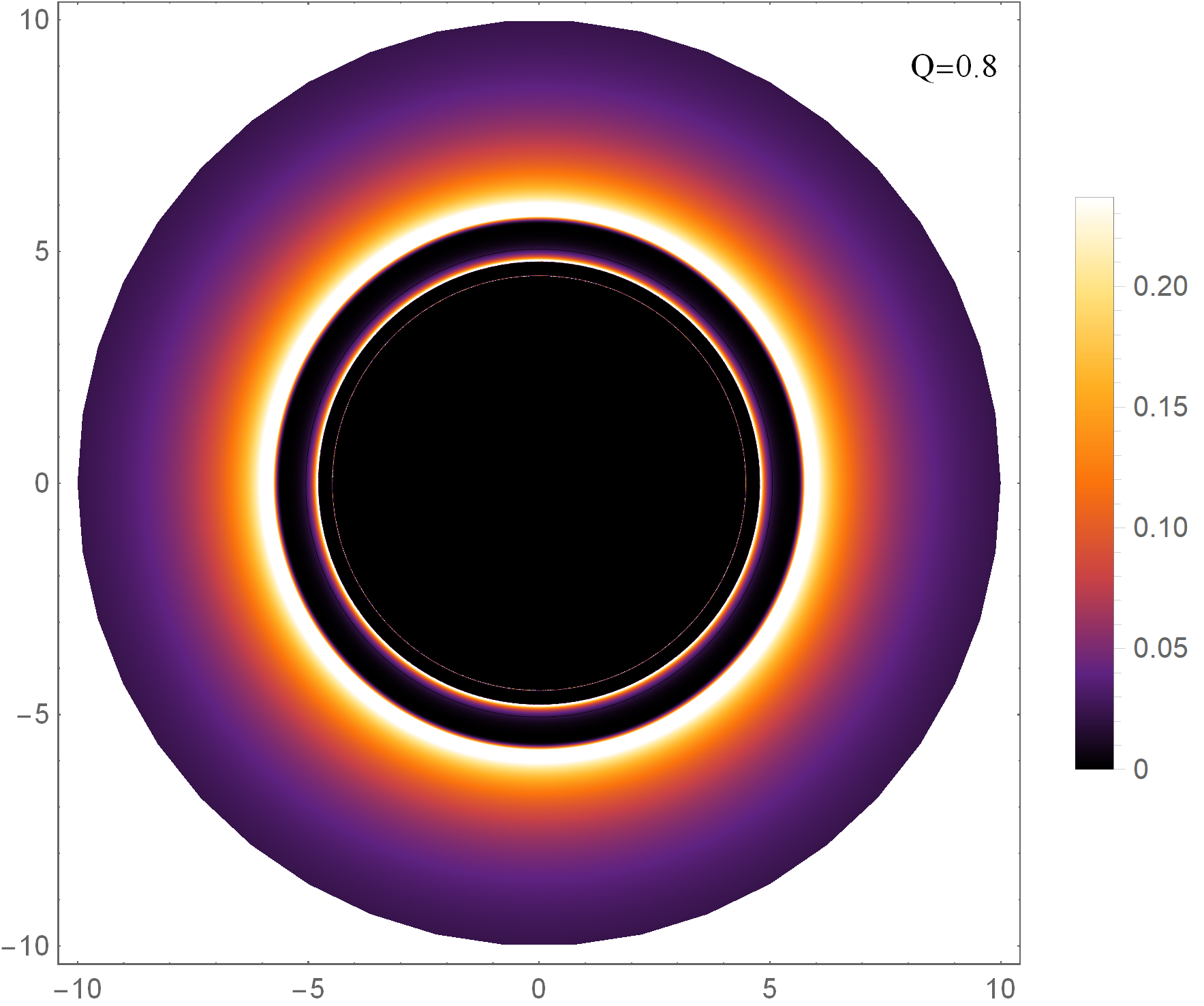}
	\includegraphics[width=5.9cm,height=5.0cm]{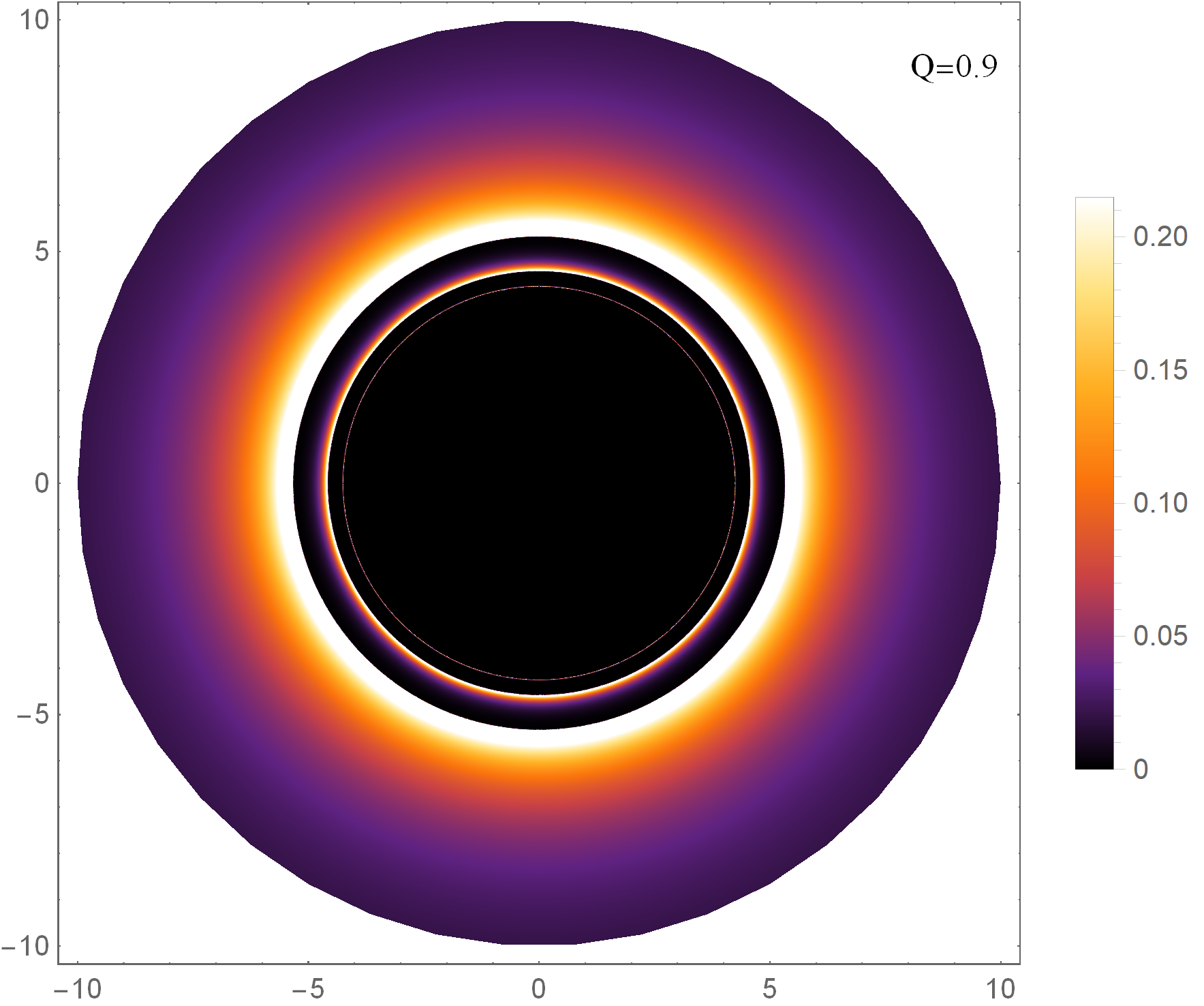}
	\caption{The observational appearance of emission near the charged black-bounce surrounded by a thin disk accretion for the choices of the bounce parameter $a=0.5$ and the charge $Q=0.1, 0.2, ..., 0.9$, respectively.}
	\label{fig:thin1}
\end{figure*}

Moreover, the observational appearance is also affected by the charge of black-bounces. In the appearance image shown in Fig.~\ref{fig:thin}, we already know that the bounce parameters have no observable effects on the rings and shadows. The appearance is the same for different values of bounce parameters. In the case of the same bounce parameter but different charges, from $0.1$ to $0.9$,  we can observe the effect of charges on the shadow and appearance in more detail in Fig.~\ref{fig:thin1}. That is, there are two remarkable features in its appearance. One is that the increasing in charge leads to the decreasing of the radii of photon rings and shadows, and the other is that the increasing in charge leads to the increasing of the observed intensity, which is manifested in the appearance of brighter and wider rings.

\subsection{Spherically symmetric infalling accretions}
   In this subsection we  investigate a spherically symmetric free-falling accretion. For this  model the intersections of light rays and the accretion appear in the whole space surrounded by the  spherical accretion. The observed intensity at the photon frequency $\nu_{\rm obs}$ can be obtained \cite{Bambi:2013nla} by integrating along the photon path with the impact parameter $b_\gamma$,
 \begin{eqnarray}
	I(\nu_{\rm obs},b_\gamma) = \int_\gamma g^3 j(\nu_e) dl_{\rm prop},
	\label{eq:intensitysp}
\end{eqnarray}
     where  $j(\nu_e)$ is the emissivity per unit volume, $\nu_e$ is the photon frequency, and $dl_{\rm prop}$ is the infinitesimal proper length in the rest frame of emitter. The redshift is given by
\begin{eqnarray}
	g = \frac{k_\mu u^\mu_{\rm obs}}{k_\nu u^\nu_e},
\end{eqnarray}
where $k^\mu$ is the 4-velocity of photons and $u^\mu_{\rm obs}$ is the 4-velocity of the distant observer. In a static and spherically symmetric spacetime, the accretion falls with the 4-velocity $u^\mu_e$,
\begin{eqnarray}
	u^t_e=\frac{1}{A(x)}, \qquad u^x_e=-\sqrt{1-A(x)}, \qquad u^{\theta}_e=u^{\varphi}_e=0,
\end{eqnarray}
and the 4-velocity of photons is
\begin{eqnarray}
	k_t = \frac{1}{b}, \qquad k_x = \pm \frac{1}{b}\sqrt{\frac{1}{A(x)}\left( \frac{1}{A(x)} - \frac{b^2}{x^2}\right) }.
\end{eqnarray}
The proper distance along a photon path $\gamma$ reduces to
\begin{eqnarray}
	dl_\gamma = k_\mu u^\mu_e d\lambda = \frac{k_t}{g |k_x|}dx.
\end{eqnarray}
For a simple model of the monochromatic emission, the specific emissivity   can be determined~\cite{Bambi:2013nla,Saurabh:2020zqg,Zeng:2020dco,Qin:2020xzu} by the delta function with a $1/x^2$ radial profile,
\begin{eqnarray}
	j(\nu_e) \propto \frac{\delta(\nu_e - \nu_*)}{x^2},
\end{eqnarray}
where $\nu_*$ is the emitted frequency in the rest frame. Then  integrating Eq.~(\ref{eq:intensitysp})  over all the observed frequencies, we
derive the total observed photon flux,
 \begin{eqnarray}
	F_{\rm obs}(b_\gamma) \propto \int_\gamma \frac{g^3}{x^2} \frac{k_t}{|k_x|} dx.
\end{eqnarray}

\begin{figure*}[htbp]
	\centering
	\includegraphics[width=6.9cm,height=5.6cm]{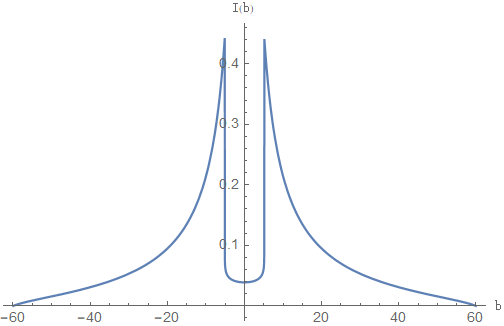}
	\includegraphics[width=6.9cm]{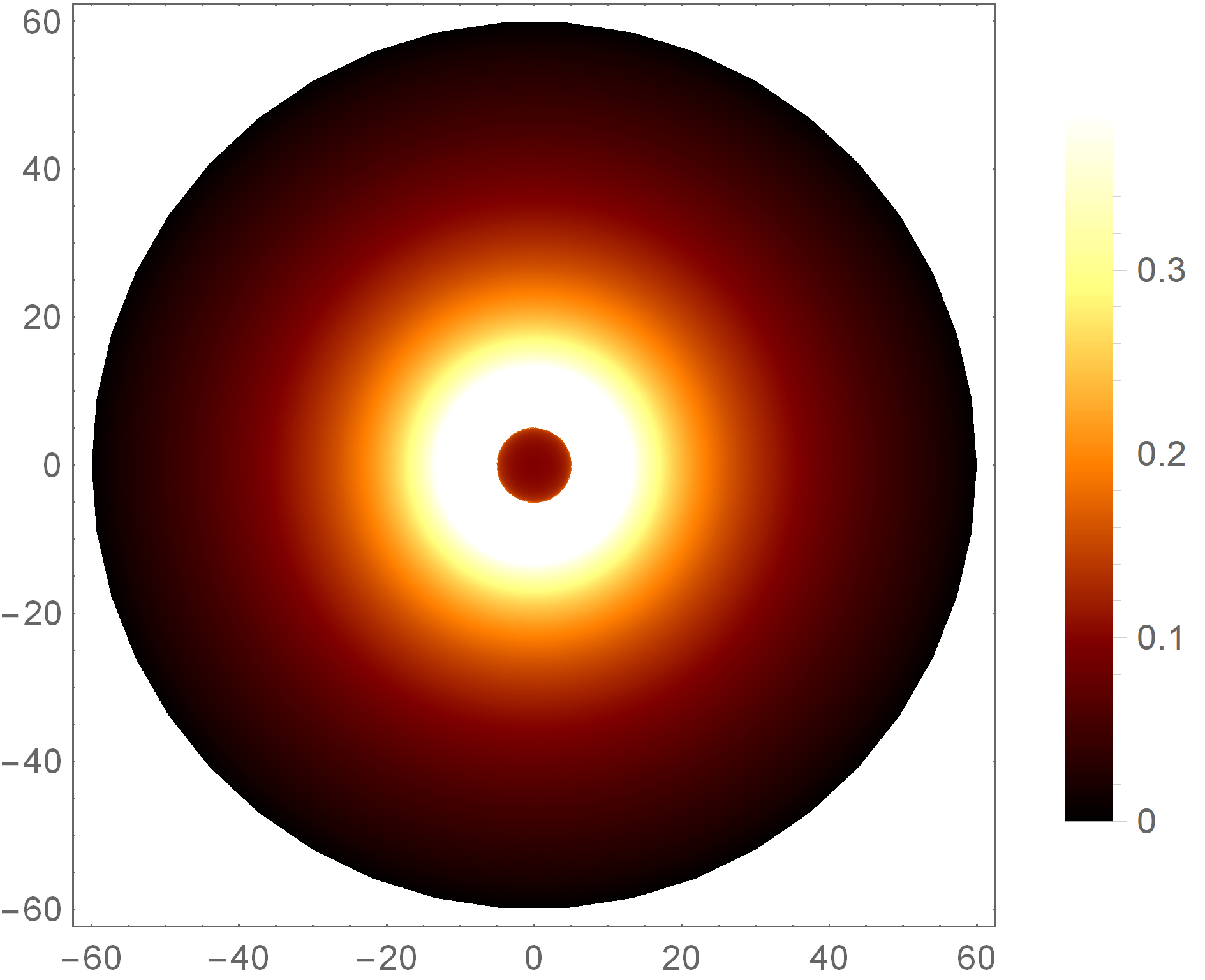}
	\includegraphics[width=6.9cm,height=5.6cm]{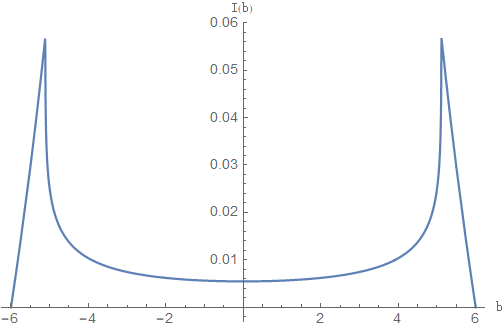}
	\includegraphics[width=6.9cm]{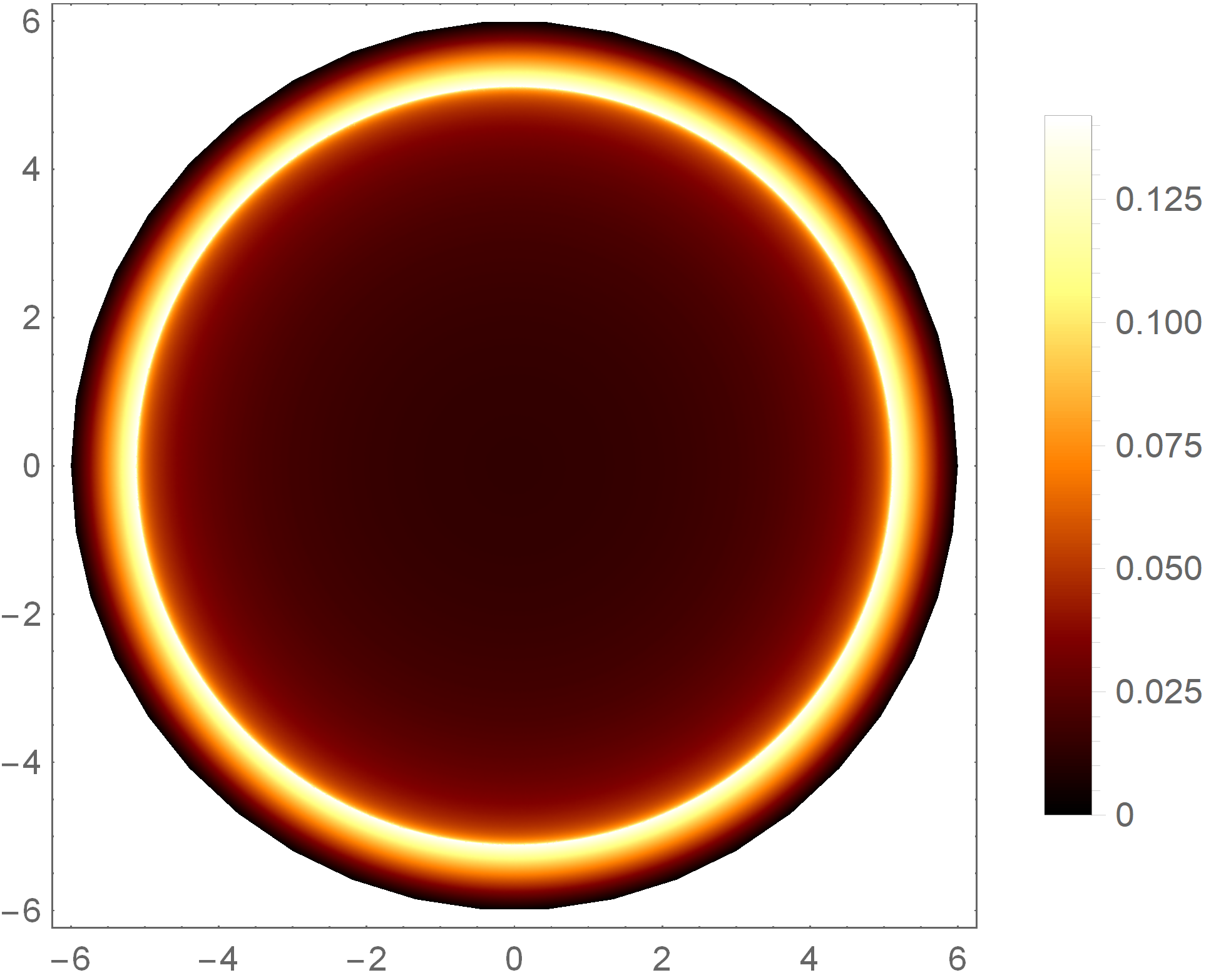}
	\caption{The observational appearance of emission near the charged black-bounce with $M=1$, $Q=0.1$ and $a=0.5$ surrounded by a spherically symmetric infalling accretion. From left to right, the panels show the observed intensity and optical appearance, respectively. The bottom row of the figure shows an image with 1/10 of the field of view of the top row.} \label{SphericalOB0.1Q}
\end{figure*}

\begin{figure*}[htbp]
	\centering
	\includegraphics[width=6.9cm,height=5.6cm]{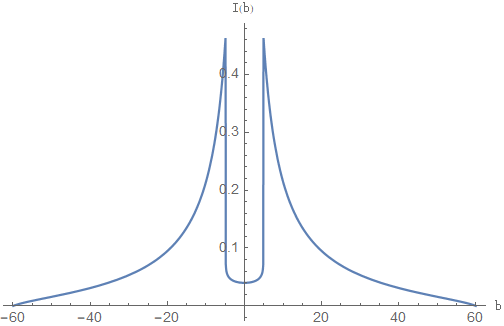}
	\includegraphics[width=6.9cm]{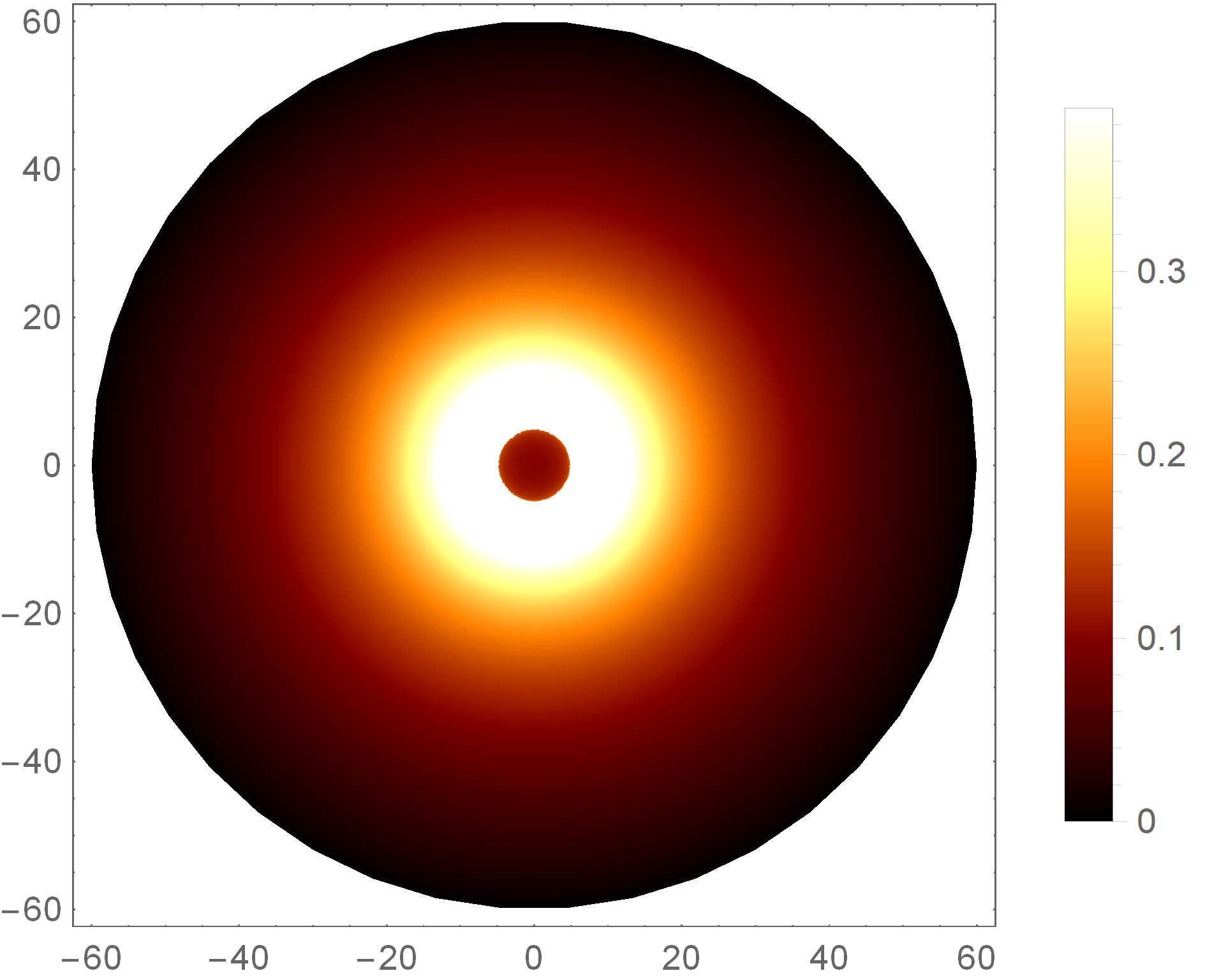}
	\includegraphics[width=6.9cm,height=5.6cm]{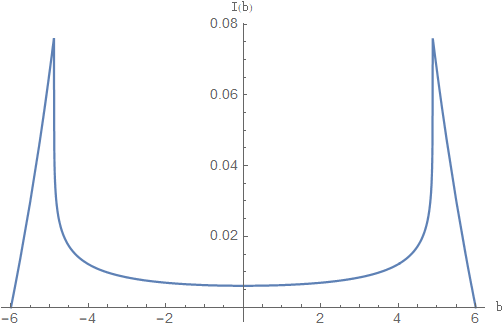}
	\includegraphics[width=6.9cm]{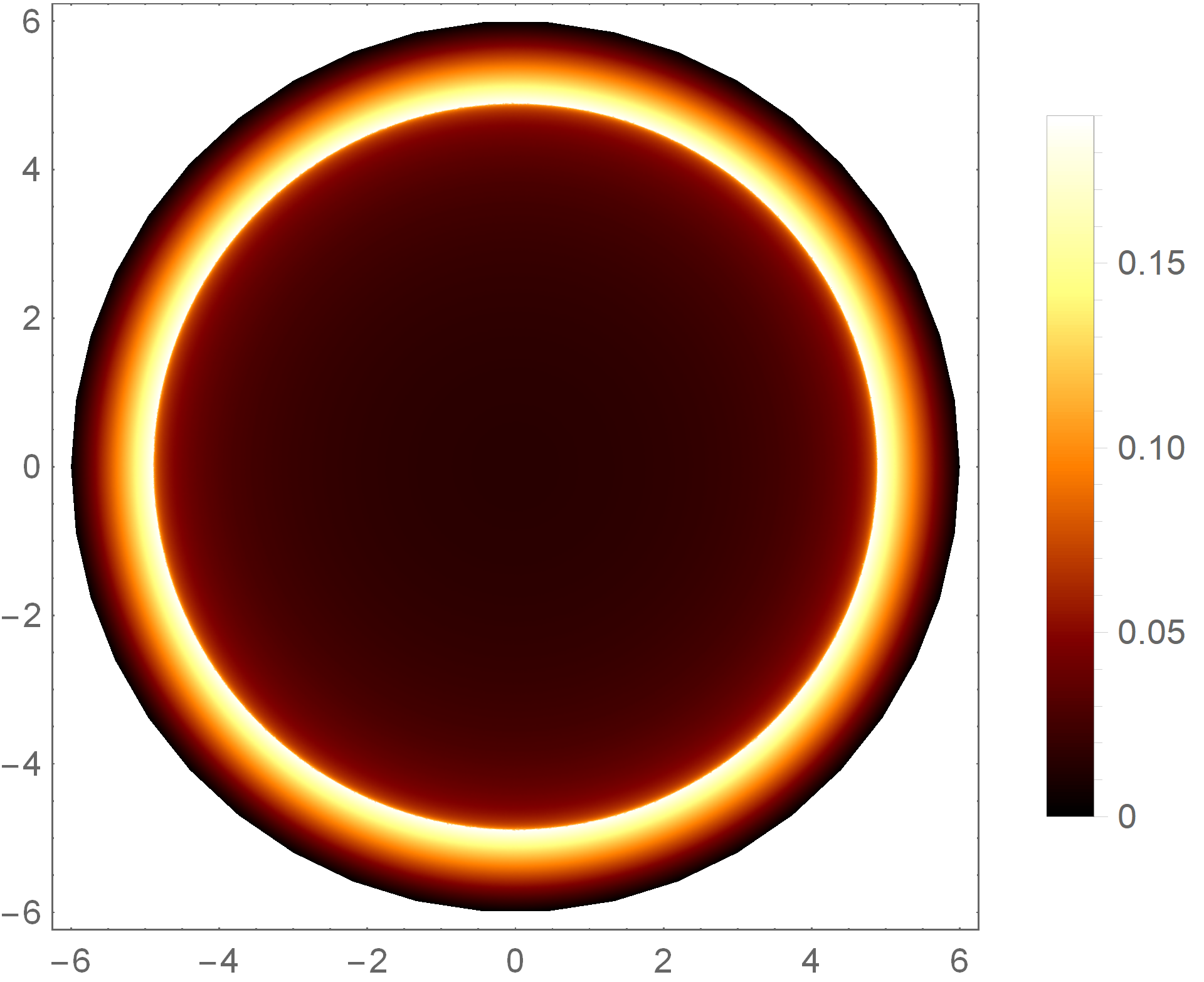}
	\caption{The observational appearance of emission near the charged black-bounce with $M=1$, $Q=0.5$ and $a=0.5$ surrounded by a spherically symmetric infalling accretion. From left to right, the panels show the observed intensity and optical appearance, respectively. The bottom row of the figure shows an image with 1/10 of the field of view of the top row.} \label{SphericalOB0.5Q}
\end{figure*}

The observed intensity and the observational appearance of images for a charged black-bounce are shown in Figs.~\ref{SphericalOB0.1Q} and \ref{SphericalOB0.5Q}, where $M=1$, $Q=0.1$ and $a=0.5$ are set in Fig.~\ref{SphericalOB0.1Q}, and $M=1$, $Q=0.5$ and $a=0.5$ are set in Fig.~\ref{SphericalOB0.5Q}.
As in the thin disk accretion, there is a peak in the observed intensity of a spherically symmetric infalling accretion (see the top left panels of Figs.~\ref{SphericalOB0.1Q} and \ref{SphericalOB0.5Q}). With the increasing of the impact parameter, the observed intensity peaks sharply at $b=b_c$ and then gradually decreases. If comparing a spherically symmetric infalling accretion with a thin disk accretion, we find that the former has a very wide luminosity which can well be presented in the optical appearance. We can clearly see from the top right panels of Figs.~\ref{SphericalOB0.1Q} and \ref{SphericalOB0.5Q} that the black circle at the center is surrounded by a very thick and bright appearance.

In the bottom rows of Figs.~\ref{SphericalOB0.1Q} and \ref{SphericalOB0.5Q} we take the field of view to 1/10 of the images in the top rows for a clear view of the photon ring.
The photon ring located at $b\approx5.19$ in the case of $Q=0.1$ is shown in Fig.~\ref{SphericalOB0.1Q} and at $b\approx4.97$ in the case of $Q=0.5$ is shown in Fig.~\ref{SphericalOB0.5Q}. Furthermore, a large charge increases the intensity of incoming light but decreases the apparent size of the shadow. The peak of the observed intensity in the case of $Q=0.5$ is higher than that in the case of $Q=0.1$, see the bottom left panels in Figs.~\ref{SphericalOB0.1Q} and \ref{SphericalOB0.5Q}. The apparent size of shadows is reduced from 5.19 to 4.97 with the reduction of charge from 0.5 to 0.1. The difference of shadows is not obvious in the top rows of Figs.~\ref{SphericalOB0.1Q} and \ref{SphericalOB0.5Q} because the optical appearances are generated in a large field of view, but such a difference can be clearly seen if we look at the images in 1/10 of the field of view, see the bottom rows. The features of photon rings in a spherically symmetric infalling accretion are same as those in a thin disk accretion discussed in the previous subsection, so that the photon ring is independent of the types of the two accretions.

\section{Discussions and comparisons with astronomical observations}
\label{sec:compare}
\subsection{Constraints on the shadow size and charge from the EHT  observations}
The variation of shadow radii for a nonrorating object with one (electric) charge or for a rotating object with two (one electric and the other angular momentum) charges has been studied~\cite{EventHorizonTelescope:2021dqv}, where the electric charge gives the same constraint on shadows as the angular momentum does in the Kerr black hole, i.e., the increasing of charges reduces the apparent size of the shadow. This is clearly consistent with our results.

On the other hand, the charge is constrained by the reconstructed shadow size of the latest EHT observations.
For RN black holes, the bound from the EHT M87$^*$ observations was given~\cite{EventHorizonTelescope:2021dqv} by
\begin{eqnarray}
	0<Q\lesssim0.9.
\end{eqnarray}
Recently, the different charge bounds have been reported~\cite{Vagnozzi:2022moj} in different types of geometries, such as the singular black holes, regular black holes and wormholes. We emphasize that
the RN-SV solution is compatible with the various charge bounds in both the theoretical and observational aspects. One reason is that the bounce parameter imposes no observable effects on the photon ring and the other reason is that the RN-SV solution shares the same critical impact parameter theoretically as the RN solution does.
\subsection{Photon rings and shadows of accreting black holes}

The question, whether the size of photon rings and the size of shadows would be affected by accretion details, has been debated.
It was claimed~\cite{Gralla:2019xty} that the size of shadows depends on emission details. However, for a simple model --- the Schwarzschild black hole surrounded by a thin disk accretion, it was shown~\cite{Narayan:2019imo} that the size of shadows is an intrinsic signature of the spacetime geometry and it is hardly affected by accretions. That is, Ref.~\cite{Narayan:2019imo} provides one counterexample to Ref.~\cite{Gralla:2019xty}. We noticed in Sec. IV that the photon ring remains unchanged in the thin disk and spherical accretions,  which provides the evidence that the photon ring is an intrinsic property of spacetime geometry. 
In addition, one recent work \cite{Chael:2021rjo} supports our opinion and that of Ref.~\cite{Gralla:2019xty}, where the photon ring is an intrinsic property but the size of shadows depends on the details of the emission region for a thin disk accretion.

%Moreover, real hot accretion flows occur~\cite{Yuan:2014gma} at low mass accretion rates and the radiative efficiency of these flows decreases with the decreasing of the mass accretion rate. Moreover, why are accretion disks associated with jets and what role does the spin of black holes play in determining jets remain unanswered, where the spin complicates accretion models. %Although real black holes are not expected to be spherically symmetric, the spin complicates the accretion models. 
%These uncertainties bring great difficulties to the detailed modeling of real accretion disks. Fortunately, the simplified models, such as the two models discussed in Sec.~\ref{sec:appearance}, capture the key features of GRMHD models, where one feature is that the emissivity increases almost monotonically inward and toward the horizon (see the left panels of Fig.~\ref{fig:thin} and Fig. 5 in Ref.~\cite{Narayan:2019imo}), and the other feature is that the clear shadow region distinguishes the charged black-bounce with spin from the RN black hole without spin.

%\subparagraph{Comments on RN-SV model}Shadow  phenomenon  contains observable information from all metric theories of gravity and can be used to test their consistency with astronomical observations

\section{Concluding remarks}
	\label{sec:con}
In the present work we have studied the photon ring, shadow and optical appearance of a charged black-bounce under various illumination conditions. The introduction of a charge term in a black-bounce spacetime  decreases the critical impact parameter, but the introduction of a length scale in the regularizing procedure does not change the critical impact parameter. We determine the upper and lower limits of  the critical impact parameter on the premise of reserving the regularity in charged black-bounces  interpolating between regular black holes and charged traversable wormholes. We also notice that the charged black-bounces and RN black holes share the same critical impact parameter. The relations between the impact parameter and charge can be reflected more clearly in the observational appearance of the region near a charged black-bounce when there are emissions from a thin disk accretion and a spherically symmetric infalling accretion. We find that a large charge increases the intensity of incoming light but decreases the apparent size of  shadows, and that the photon rings remain unchanged in the two different accretion models. These results are consistent with the recent observations,  which confirms that the photon ring is the intrinsic property of a spacetime geometry.

\section*{Acknowledgments}
The authors would like to thank the anonymous referee for the helpful comments that improve
this work greatly. This work was supported in part by the National Natural Science Foundation of China under Grant Nos. 11675081 and 12175108.

\bibliography{references}
\end{document}